\documentclass[a4paper,12pt]{article}
\usepackage{graphicx}
\usepackage{ifthen} % for conditional statements

\newboolean{articletitles}
\setboolean{articletitles}{true} % False removes titles in references

\newboolean{uprightparticles}
\setboolean{uprightparticles}{false} %True for upright particle symbols
\usepackage{amsmath} % Adds a large collection of math symbols
\usepackage{amssymb}
\usepackage{amsfonts}
\usepackage{upgreek} % Adds in support for greek letters in roman typeset
\usepackage{xspace} 
\usepackage{upgreek}

\def\lhcb {\mbox{LHCb}\xspace}
\def\atlas  {\mbox{ATLAS}\xspace}
\def\cms    {\mbox{CMS}\xspace}

\def\babar  {\mbox{BaBar}\xspace}
\def\belle  {\mbox{Belle}\xspace}

\def\MagUp {\mbox{\em Mag\kern -0.05em Up}\xspace}

\ifthenelse{\boolean{uprightparticles}}%
{

 \def\Pmu         {\ensuremath{\upmu}\xspace}                 
 \def\Pnu         {\ensuremath{\upnu}\xspace}                 
                  
 \def\Ppi         {\ensuremath{\uppi}\xspace}

 \def\Ptau        {\ensuremath{\uptau}\xspace}

 \def\Ppsi        {\ensuremath{\uppsi}\xspace}

 \def\PDelta      {\ensuremath{\Delta}\xspace}                 
 \def\PXi      {\ensuremath{\Xi}\xspace}                 
 \def\PLambda      {\ensuremath{\Lambda}\xspace}                 
 \def\PSigma      {\ensuremath{\Sigma}\xspace}                 
 \def\POmega      {\ensuremath{\Omega}\xspace}                 
 \def\PUpsilon      {\ensuremath{\Upsilon}\xspace}                 
 
 \def\PB      {\ensuremath{\mathrm{B}}\xspace}                 
                  
 \def\PD      {\ensuremath{\mathrm{D}}\xspace}

 \def\PJ      {\ensuremath{\mathrm{J}}\xspace}                 
 \def\PK      {\ensuremath{\mathrm{K}}\xspace}

 \def\Pb      {\ensuremath{\mathrm{b}}\xspace}                 
 \def\Pc      {\ensuremath{\mathrm{c}}\xspace}

 \def\Pi      {\ensuremath{\mathrm{i}}\xspace}

 \def\Pp      {\ensuremath{\mathrm{p}}\xspace}

 \def\Ps      {\ensuremath{\mathrm{s}}\xspace}

}
{

 \def\Pmu         {\ensuremath{\mu}\xspace}                 
 \def\Pnu         {\ensuremath{\nu}\xspace}                 
                  
 \def\Ppi         {\ensuremath{\pi}\xspace}

 \def\Ptau        {\ensuremath{\tau}\xspace}

 \def\Ppsi        {\ensuremath{\psi}\xspace}                 
                  
 \mathchardef\PDelta="7101
 \mathchardef\PXi="7104
 \mathchardef\PLambda="7103
 \mathchardef\PSigma="7106
 \mathchardef\POmega="710A
 \mathchardef\PUpsilon="7107
                  
 \def\PB      {\ensuremath{B}\xspace}                 
                  
 \def\PD      {\ensuremath{D}\xspace}

 \def\PJ      {\ensuremath{J}\xspace}                 
 \def\PK      {\ensuremath{K}\xspace}

 \def\Pb      {\ensuremath{b}\xspace}                 
 \def\Pc      {\ensuremath{c}\xspace}

 \def\Pi      {\ensuremath{i}\xspace}

 \def\Pp      {\ensuremath{p}\xspace}

 \def\Ps      {\ensuremath{s}\xspace}

}

\makeatletter
\ifcase \@ptsize \relax% 10pt
  \newcommand{\miniscule}{\@setfontsize\miniscule{4}{5}}% \tiny: 5/6
\or% 11pt
  \newcommand{\miniscule}{\@setfontsize\miniscule{5}{6}}% \tiny: 6/7
\or% 12pt
  \newcommand{\miniscule}{\@setfontsize\miniscule{5}{6}}% \tiny: 6/7
\fi
\makeatother

\DeclareRobustCommand{\optbar}[1]{\shortstack{{\miniscule (\rule[.5ex]{1.25em}{.18mm})}
  \\ [-.7ex] $#1$}}

\def\mup        {{\ensuremath{\Pmu^+}}\xspace}
\def\mun        {{\ensuremath{\Pmu^-}}\xspace} % muon negative (\mum is taken)
\def\mumu       {{\ensuremath{\Pmu^+\Pmu^-}}\xspace}

\def\taup       {{\ensuremath{\Ptau^+}}\xspace}
\def\taum       {{\ensuremath{\Ptau^-}}\xspace}

\def\neu        {{\ensuremath{\Pnu}}\xspace}
\def\neub       {{\ensuremath{\overline{\Pnu}}}\xspace}
\def\squark    {{\ensuremath{\Ps}}\xspace}

\def\cquark    {{\ensuremath{\Pc}}\xspace}

\def\bquark    {{\ensuremath{\Pb}}\xspace}

\def\pion   {{\ensuremath{\Ppi}}\xspace}
\def\piz    {{\ensuremath{\pion^0}}\xspace}

\def\pip    {{\ensuremath{\pion^+}}\xspace}
\def\pim    {{\ensuremath{\pion^-}}\xspace}

\def\kaon    {{\ensuremath{\PK}}\xspace}
  \def\Kbar    {{\kern 0.2em\overline{\kern -0.2em \PK}{}}\xspace}

\def\KorKbar    {\kern 0.18em\optbar{\kern -0.18em K}{}\xspace}
\def\Kz      {{\ensuremath{\kaon^0}}\xspace}

\def\Kp      {{\ensuremath{\kaon^+}}\xspace}

\def\KS      {{\ensuremath{\kaon^0_{\mathrm{ \scriptscriptstyle S}}}}\xspace}

\def\Kstarz  {{\ensuremath{\kaon^{*0}}}\xspace}
\def\Kstarzb {{\ensuremath{\Kbar{}^{*0}}}\xspace}
\def\Kstar   {{\ensuremath{\kaon^*}}\xspace}

\def\Kstarp  {{\ensuremath{\kaon^{*+}}}\xspace}

  \def\Dbar    {{\kern 0.2em\overline{\kern -0.2em \PD}{}}\xspace}
\def\D       {{\ensuremath{\PD}}\xspace}

\def\DorDbar    {\kern 0.18em\optbar{\kern -0.18em D}{}\xspace}
\def\Dz      {{\ensuremath{\D^0}}\xspace}

\def\Dp      {{\ensuremath{\D^+}}\xspace}

\def\Dsp     {{\ensuremath{\D^+_\squark}}\xspace}

\def\B       {{\ensuremath{\PB}}\xspace}
\def\Bbar    {{\ensuremath{\kern 0.18em\overline{\kern -0.18em \PB}{}}}\xspace}

\def\BorBbar    {\kern 0.18em\optbar{\kern -0.18em B}{}\xspace}

\def\Bu      {{\ensuremath{\B^+}}\xspace}
\def\Bub     {{\ensuremath{\B^-}}\xspace}

\def\Bd      {{\ensuremath{\B^0}}\xspace}
\def\Bs      {{\ensuremath{\B^0_\squark}}\xspace}

\def\jpsi     {{\ensuremath{{\PJ\mskip -3mu/\mskip -2mu\Ppsi\mskip 2mu}}}\xspace}

  \def\Y#1S{\ensuremath{\PUpsilon{(#1S)}}\xspace}% no space before {...}!

\def\proton      {{\ensuremath{\Pp}}\xspace}
\def\antiproton  {{\ensuremath{\overline \proton}}\xspace}

\def\Lz          {{\ensuremath{\PLambda}}\xspace}
\def\Lbar        {{\ensuremath{\kern 0.1em\overline{\kern -0.1em\PLambda}}}\xspace}
\def\LorLbar    {\kern 0.18em\optbar{\kern -0.18em \PLambda}{}\xspace}
\def\Lambdares   {{\ensuremath{\PLambda}}\xspace}

\def\Sigmares    {{\ensuremath{\PSigma}}\xspace}

\def\Lb      {{\ensuremath{\Lz^0_\bquark}}\xspace}

\def\Lc      {{\ensuremath{\Lz^+_\cquark}}\xspace}

\def\BF         {{\ensuremath{\mathcal{B}}}}

\def\BR         {\BF}
\newcommand{\decay}[2]{\ensuremath{#1\!\to #2}\xspace}         % {\Pa}{\Pb \Pc}

\def\to                 {\ensuremath{\rightarrow}\xspace}

\def\qsq       {{\ensuremath{q^2}}\xspace}

\def\CP                {{\ensuremath{C\!P}}\xspace}

\def\BdToKstmm    {\decay{\Bd}{\Kstarz\mup\mun}}

\def\AT#1     {\ensuremath{A_{\mathrm{T}}^{#1}}\xspace}           % 2

\def\C#1      {\ensuremath{\mathcal{C}_{#1}}\xspace}                       % 9
\def\Cp#1     {\ensuremath{\mathcal{C}_{#1}^{'}}\xspace}                    % 7
\def\Ceff#1   {\ensuremath{\mathcal{C}_{#1}^{\mathrm{(eff)}}}\xspace}        % 9  
\def\Cpeff#1  {\ensuremath{\mathcal{C}_{#1}^{'\mathrm{(eff)}}}\xspace}       % 7
\def\Ope#1    {\ensuremath{\mathcal{O}_{#1}}\xspace}                       % 2
\def\Opep#1   {\ensuremath{\mathcal{O}_{#1}^{'}}\xspace}                    % 7

\newcommand{\tev}{\ifthenelse{\boolean{inbibliography}}{\ensuremath{~T\kern -0.05em eV}}{\ensuremath{\mathrm{\,Te\kern -0.1em V}}}\xspace}
\newcommand{\gev}{\ensuremath{\mathrm{\,Ge\kern -0.1em V}}\xspace}
\newcommand{\mev}{\ensuremath{\mathrm{\,Me\kern -0.1em V}}\xspace}
\newcommand{\kev}{\ensuremath{\mathrm{\,ke\kern -0.1em V}}\xspace}
\newcommand{\ev}{\ensuremath{\mathrm{\,e\kern -0.1em V}}\xspace}
\newcommand{\gevc}{\ensuremath{{\mathrm{\,Ge\kern -0.1em V\!/}c}}\xspace}
\newcommand{\mevc}{\ensuremath{{\mathrm{\,Me\kern -0.1em V\!/}c}}\xspace}
\newcommand{\gevcc}{\ensuremath{{\mathrm{\,Ge\kern -0.1em V\!/}c^2}}\xspace}
\newcommand{\gevgevcccc}{\ensuremath{{\mathrm{\,Ge\kern -0.1em V^2\!/}c^4}}\xspace}
\newcommand{\mevcc}{\ensuremath{{\mathrm{\,Me\kern -0.1em V\!/}c^2}}\xspace}

\def\ps   {\ensuremath{{\mathrm{ \,ps}}}\xspace}

\def\gsim{{~\raise.15em\hbox{$>$}\kern-.85em
          \lower.35em\hbox{$\sim$}~}\xspace}
\def\lsim{{~\raise.15em\hbox{$<$}\kern-.85em
          \lower.35em\hbox{$\sim$}~}\xspace}

\def\tell1  {TELL1\xspace}
\def\ukl1   {UKL1\xspace}

\newcommand{\ie}{\mbox{\itshape i.e.}\xspace}

\usepackage{cite} % Allows for ranges in citations
\usepackage{mciteplus}

\begin{document}

\begin{flushright}
\today
\end{flushright}

\vspace{0.3cm}

\begin{center}
{\bf {\Large Proceedings of the Working group 3\\
``Rare B, D and K decays''\\[0.2cm]
for CKM 2018 (Heidelberg)}}

\vspace{0.5cm}

Michel De Cian$^{a}$, S\'ebastien Descotes-Genon$^{b}$, Karim Massri$^{c}$

\vspace{0.2cm}

\emph{$^a$ Laboratoire de Physique des Hautes \'Energies, EPFL, 1015 Lausanne, Switzerland}\\
\emph{$^b$  Laboratoire de Physique Th\'eorique (UMR 8627),\\
CNRS, Univ. Paris-Sud, Universit\'e Paris-Saclay, 91405 Orsay, France}\\
\emph{$^c$ CERN, European Organization for Nuclear Research, 1211 Geneva, Switzerland}\\

\end{center}

\vspace{0.5cm}

\emph{We provide an overview of the results presented during the sessions of Working Group 3 ``Rare B, D and K decays, radiative and electroweak penguin decays,
including constraints on $V_{td}/V_{ts}$ and $\epsilon'/\epsilon$", presented at the 10th International Workshop on the CKM Unitarity Triangle (CKM 2018) at Heidelberg University (September 17-21, 2018).}

\vspace{0.5cm}

%\section{Introduction}

Rare \B, \D and \kaon decays provide interesting probes of the Standard Model (SM), with a potential sensitivity to New Physics (NP) higher than other, more common, decays. Their experimental measurement is challenging, and their theoretical interpretation requires a precise knowledge of QCD at low energies, in the non-perturbative (hadronic) regime. The various \bquark, \cquark, \squark decays provide different types of tests of the Standard Model, which can be performed in different experimental settings and which may exhibit correlated deviations in models of New Physics.

\section{$B$ decays}

\subsection{Exclusive $b \to s \ell \ell$ transitions}
$b \to s \ell \ell$ transitions are an excellent probe for physics beyond the SM, as they are forbidden on tree level. Several deviations from SM predictions have been observed in the differential branching ratios, the angular distributions and the ratio of the branching ratios of decays with muon and electron final states.

\subsubsection{Differential branching ratios}
The differential branching ratios of the decays \BdToKstmm\cite{Aaij:2016flj}, \decay{\Bu}{\Kstarp}{\mumu}, \decay{\Bu}{\Kp\mumu}, \decay{\Bd}{\Kz\mumu}\cite{Aaij:2014pli}, \decay{\Bs}{\phi\mumu}\cite{Aaij:2015esa} and \decay{\Lb}{\Lambdares\mumu}\cite{Aaij:2015xza} have been measured by the \lhcb experiment. All measurements show smaller values than predicted in the region of low dilepton invariant mass, \qsq, albeit large uncertainties in the theoretical predictions are present and limit the precision of these measurements.

\subsubsection{Angular distributions}
In the decays \BdToKstmm\cite{Aaij:2015oid, Sirunyan:2017dhj, Khachatryan:2015isa, Aaboud:2018krd}, \decay{\Bu}{\Kp\mumu}\cite{CMS:2018sus} and \decay{\Lb}{\Lambdares\mumu}\cite{Aaij:2018gwm} the angular distributions have been measured in the full \qsq region ($\BdToKstmm$, \decay{\Bu}{\Kp\mumu} ) and in the \qsq interval 15\gevgevcccc -- 20\gevgevcccc (\decay{\Lb}{\Lambdares\mumu}).

In the decay \decay{\Lb}{\Lambdares\mumu} for the first time all 34 angular observables were measured using a moments analysis. All results are in agreement with their SM predictions. The decay \decay{\Bu}{\Kp\mumu} has only two angular observables, the forward-backward asymmetry $A_{FB}$ and $F_{H}$, sensitive to (pseudo-)scalar and tensor contributions. Both observables were measured to be compatible with SM predictions.

In the decay \BdToKstmm, measurements of all angular parameters were performed, with the observable $P_{5}^{'}$ showing a deviation from the SM by 3.4 standard deviations in the \lhcb measurement. A deviation of $P_{5}^{'}$ could be due to a value of the Wilson coefficient \C9, describing the strength of the vector coupling, different from its SM prediction. Or the discrepancy could be due to an underestimation of charm-loop effects. A possible strategy to distinguish both contributions is to fit the differential branching fraction of \BdToKstmm, modelling the contributing tree and penguin amplitudes using Breit-Wigner line shapes, to determine the relative contribution of long-distance (\ie charm loop) and short-distance contributions. This measurement was already performed for the decay \decay{\Bu}{\Kp\mumu}\cite{Aaij:2016cbx}. An alternative approach is to fit the continuous distribution of $P_{5}^{'}$ as described in Refs.~\cite{Blake:2017fyh} and \cite{Bobeth:2017vxj}.

\subsubsection{Lepton Flavour Universality violation}
The ratios $R_{K} = \frac{ \BR(\decay{\Bu}{\Kp\mumu}) }{\BR(\decay{\Bu}{\Kp e^{+}e^{-}})}$ and $R_{K^{*}} = \frac{ \BR(\decay{\Bd}{\Kstarz\mumu}) }{\BR(\decay{\Bd}{\Kstarz e^{+}e^{-}})}$ have been measured at the \lhcb\cite{Aaij:2014ora, Aaij:2017vbb}, \babar\cite{Lees:2012tva} and \belle\cite{Wei:2009zv} experiments. While the precision of the results from \babar and \belle are limited by the statistical precision, the \lhcb results show a deviation from the SM. $R_{K}$ is measured to be $R_{K} = 0.745^{+0.090}_{-0.074} \pm 0.036$, in the \qsq interval 1\gevgevcccc -- 6 \gevgevcccc, corresponding to a deviation of 2.6$\sigma$ from the SM.
$R_{K^{*}}$ is measured to be $R_{K^{*}} = 0.66^{+0.11}_{-0.07} \pm 0.03$ in the \qsq interval 0.045\gevgevcccc -- 1.1\gevgevcccc and  $R_{K^{*}} = 0.69^{+0.11}_{-0.07} \pm 0.05$ in the \qsq interval 1.1\gevgevcccc -- 6.0\gevgevcccc. This corresponds to a deviation of 2.2 and 2.4 standard deviations from the SM. The predictions for both measurements have a negligible theoretical uncertainty, as hadronic effects cancel when forming the ratio.

\subsubsection{$\tau$ leptons in the final state}
So far, only $\bquark \to \squark \ell \ell$ transitions have been measured where $\ell$ is a muon or an electron. Given the low branching ratio of $\mathcal{O}(10^{-7})$ and the experimental challenge to reconstruct $\tau$ leptons, $\bquark \to \squark \taup \taum$ processes are experimentally only weakly constrained. A deviation of the $R(X)$ ratios, with $X = \jpsi, \Dz^{*}, \Dp^{*}$ from the SM predictions, which is suggested by measurements of \lhcb\cite{Aaij:2017tyk, Aaij:2015yra, Aaij:2017uff}, \babar\cite{Lees:2012xj, Lees:2013uzd} and \belle\cite{Huschle:2015rga, Sato:2016svk, Hirose:2016wfn, Hirose:2017dxl}, would lead to an enhancement of the branching ratio of $\bquark \to \squark \taup \taum$ transitions, most significantly to the purely leptonic decay \decay{\Bs}{\taup\taum}. Given the possible reach with \lhcb and \belle II, a discovery of a $\bquark \to \squark \taup \taum$ processes would imply physics beyond the SM.

\subsubsection{The purely leptonic decays \decay{\Bs}{\mumu} and \decay{\Bd}{\mumu}}
\begin{figure}[h!]
  \begin{center}
\includegraphics[width=0.6\linewidth]{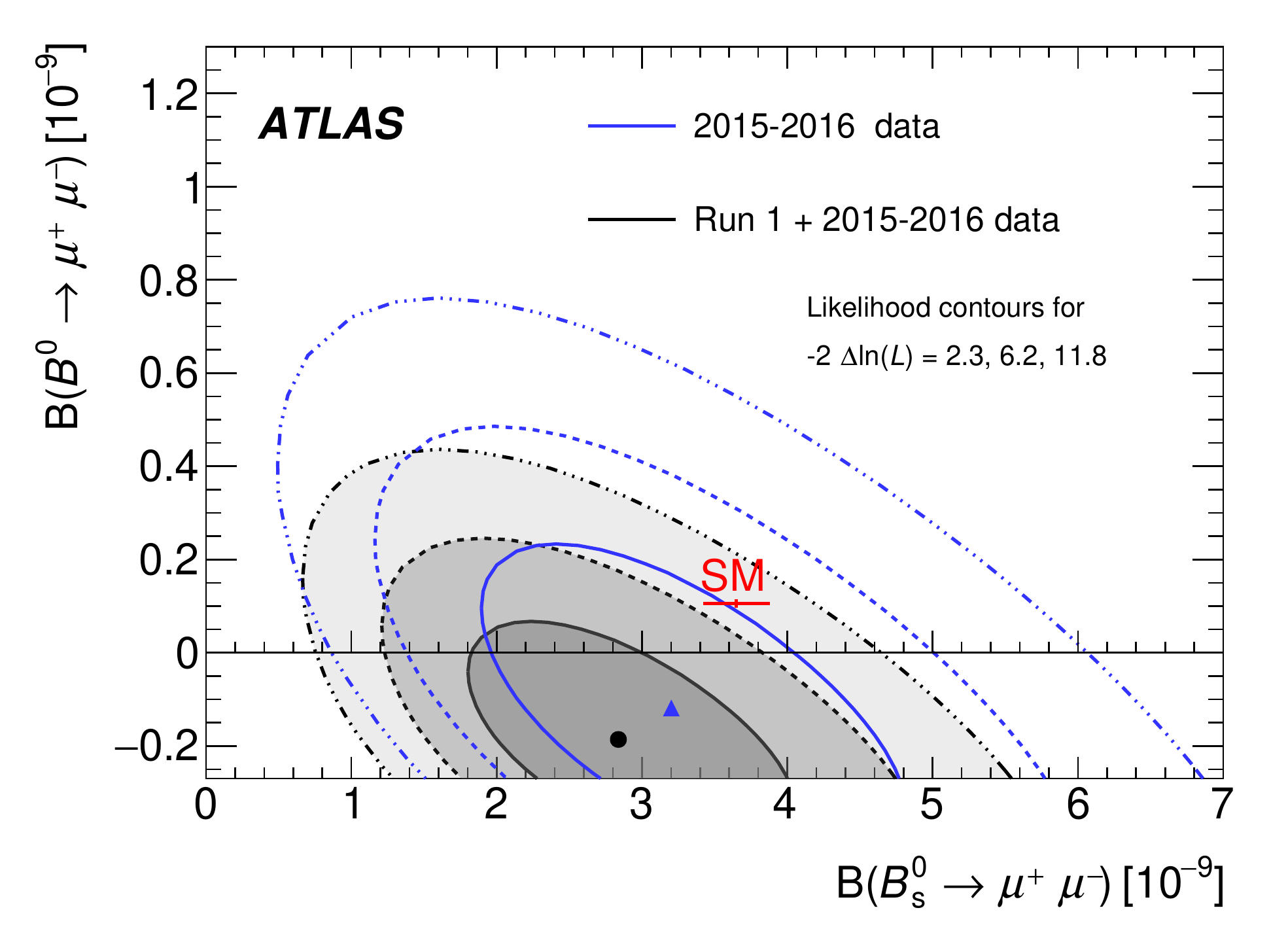}
    \vspace*{-0.5cm}
  \end{center}
  \caption{Likelihood contours for the combination of the Run 1 and 2015–-2016 Run 2 results (shaded areas). The contours are obtained from the combined likelihoods of the two analyses, for values of $-2\Delta \log(L)$ equal to 2.3, 6.2 and 11.8. The empty contours represent the result from 2015–-2016 Run 2 data alone. The SM prediction with uncertainties is indicated. Figure taken from Ref.~\cite{Aaboud:2018mst}.}
  \label{fig:ATLASBmumu}
\end{figure}

The decays \decay{\Bs}{\mumu} and \decay{\Bd}{\mumu} are strongly suppressed in the SM due to being flavour-changing neutral currents and helicity suppression. They are sensitive to the Wilson coefficient \C10 and the non-SM Wilson coefficients $\mathcal{C}_S$ and $\mathcal{C}_P$. The branching ratio of \decay{\Bs}{\mumu} has been measured by the \lhcb\cite{Aaij:2017vad}, \atlas\cite{Aaboud:2018mst} and \cms\cite{CMS:2014xfa} experiments, with the latest result from the  \atlas collaboration, being $\BR(\decay{\Bs}{\mumu}) = (2.8^{+0.8}_{-0.7})\times 10^{-9}$.
In the SM, only the heavy eigenstate decays into the dimuon final state, a measurement of the lifetime could therefore reveal a deviation from the SM. The lifetime has been measured by the \lhcb experiment to be $\tau = (2.04\pm 0.44 \pm 0.05)\ps$ and is compatible with the expectation. The precision is limited by the small number of observed events.
The decay \decay{\Bd}{\mumu} is currently unobserved and upper limits are computed, with the most stringent one by the \atlas collaboration, yielding $\BR(\decay{\Bd}{\mumu}) < 2.1 \times 10^{-10}$ at 95\% CL\cite{Aaboud:2018mst}. A plot with the \atlas results is shown in Fig.~\ref{fig:ATLASBmumu}.

\subsection{Inclusive \decay{\B}{X_{d,s}\ell\ell} decays}
$B\to X_s\ell\ell$ (and $B \to X_s\gamma$) decays provide strong constraints on the short-distance Wilson coefficients \C7 (and \C9, \C10 ), with interesting experimental prospects at Belle II. The computation relies on quark-hadron duality: the same computation can be performed using a hadronic or a quark language, the latter avoiding the complications due to hadronisation, provided that the observable is inclusive enough, \ie, summed over a large set of kinematic configurations. Experimentally, one has however to perform cuts on $E_\gamma$ or $M_{X_s}$ that are difficult to tackle, introducing (hadronic) shape functions to be modelled theoretically/fitted experimentally. The charm resonances are signalled by charm loops inducing an $m_c$ dependence,  which is a significant source of uncertainty for $B\to X_s\ell\ell$, as well as higher orders in perturbation theory~\cite{deBoer:2017way,Benzke:2017woq}. A general analytic computation for all values of $m_c$ is difficult, but it is currently under investigation, recently in the case of $B\to X_s\gamma$. Recent improvements in NNLO computations have led to SM predictions for $\B \to X_s\gamma$ with $E_\gamma>1.6$ \gev of 6.9\% (to be compared to 4.5\% for the experimental average)~\cite{Misiak:2015xwa,Czakon:2015exa}.
Another improvement in the accuracy of the prediction for $B\to X_{d/s} \ell\ell$ decays has been the recent inclusion of final states with a large number of particles, which can have a sizable contribution to inclusive $b\to s\ell\ell$ transitions~\cite{Huber:2018gii}.

\subsection{Global fits}
\begin{figure}[h!]
  \begin{center}
    \includegraphics[width=0.47\linewidth]{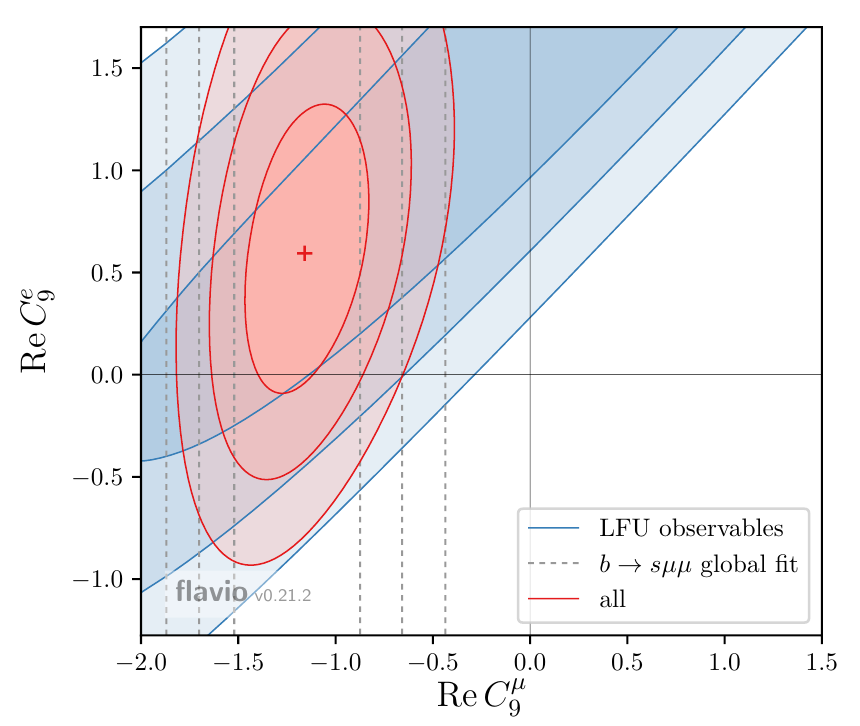}%\put(-32,133){(a)}
    \includegraphics[width=0.47\linewidth]{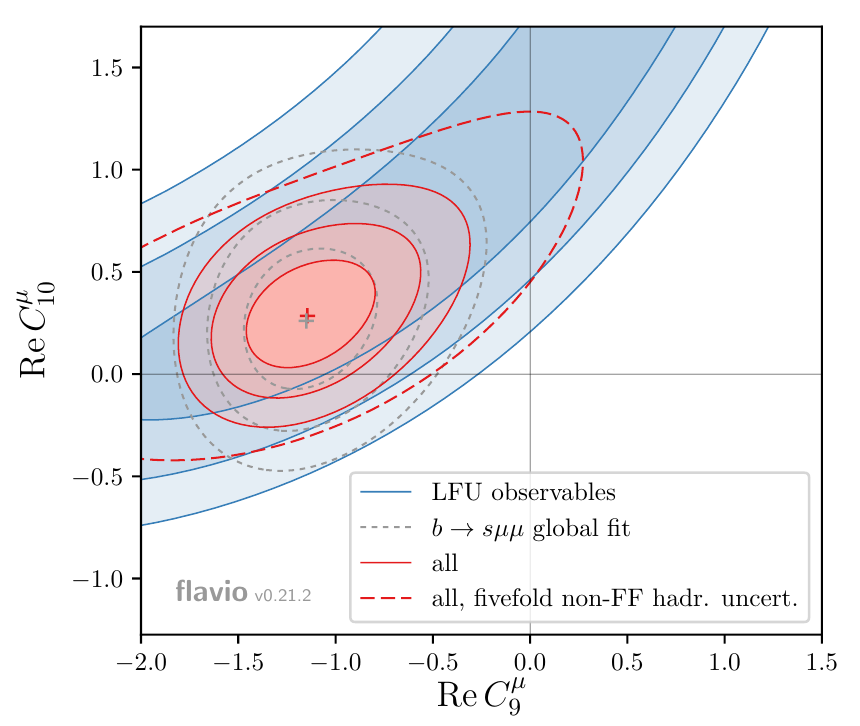}%\put(-32,133){(a)}
    \vspace*{-0.5cm}
  \end{center}
  \caption{Compatibility of the best fit of the real part of the (left) flavour-specific Wilson coefficients $\mathcal{C}_{9}^{\mu, e}$ and (right) of \C9 and \C10 with the SM prediction. Figures taken from Ref.~\cite{Altmannshofer:2017yso}.}
  \label{fig:globalFits}
\end{figure}

The measured branching ratios of the decays \decay{\B}{\Kstar \mumu}, \decay{\B}{K\mumu}, \decay{\Bs}{\phi\mumu}, the inclusive decay rate \decay{\B}{X_{s}\mumu} and the angular observables of \BdToKstmm and \decay{\Bs}{\phi\mumu} are combined in a global fit\cite{Altmannshofer:2017fio, Altmannshofer:2017yso}. The best fit point in the plane of the Wilson coefficients \C9 and \C10 shows a deviation from the SM by 5 standard deviations. When also allowing for different operators for electron or muon final states, the overall picture is unchanged, where a deviation from the SM is only seen in the muon final states, see Fig.~\ref{fig:globalFits}.

\subsection{Radiative decays}
\subsubsection{Photon polarization in \decay{\bquark}{\squark\gamma} decays}
The photon polarization in \decay{\bquark}{\squark\gamma} transitions is a sensitive probe for new physics. While it has been established that the photon is polarized in the decay \decay{\Bu}{\Kp\pim\pip\gamma}\cite{Aaij:2014wgo}, a precise determination of the polarization requires an amplitude analysis to understand the resonant structure of the \Kp\pim\pip system, and an angular analysis. Such a measurement can be expected in the near future from \belle II and \lhcb.
An alternative approach to measure the polarization is to determine the time-dependent decay rate of \decay{\B}{X_{s}\gamma} decays and the \CP parameters $C_{CP}$ and $S_{CP}$. These measurements have been performed by \belle and \babar in the decays \decay{\Bd}{\KS\piz\gamma}, \decay{\Bd}{\KS\eta\gamma} and \decay{\Bd}{\KS\eta\gamma}\cite{HFLAV16}, where all results are compatible with the SM predictions. A novel approach is to measure the decay \decay{\Bd}{\KS\pip\pim\gamma}\cite{Akar:2018zhv}, as the Dalitz structure of \KS\pip\pim allows to determine the real and imaginary part of the Wilson coefficients \C7 and $\mathcal{C}_{7}^{'}$, which is not possible when integrating over the Dalitz space.
While extracting the full information about the photon polarization from the time-dependent decay rate requires the knowledge of the flavour of the \B meson, the parameter $A^{\Delta}$, related to the polarization, can be extracted without this knowledge. This measurement was performed by \lhcb in the decay ${\decay{\Bs}{\phi\gamma}}$, yielding a value of  $A^{\Delta} = -0.98^{+0.46+0.23}_{-0.52-0.20}$\cite{Aaij:2016ofv}. This result is compatible with the SM at 2$\sigma$.

\subsubsection{Isospin and $\Delta A_{CP}$ asymmetries in \decay{B}{X_{s}\gamma} decays}

\belle has measured the isospin asymmetry and $\Delta A_{CP}$ in \decay{B}{\Kstar\gamma} decays\cite{Horiguchi:2017ntw}, where the \Kstar was reconstructed in the \Kp\pim, \Kp\piz, \KS\pip and \KS\piz final states, with event yields between about 350 to 2300 signal candidates. The \CP asymmetry was measured for the first time and yields $\Delta A_{CP} = (2.4\pm 2.8 \pm 0.5) \%$. The isospin asymmetry was measured to be $\Delta_{0+} = (6.2\pm1.5\pm0.6\pm1.2) \%$. While the \CP asymmetry is compatible with the SM prediction, the isospin asymmetry deviates by 3.1$\sigma$ from the SM prediction. 

A similar measurement was performed for the inclusive $X_{s}$ final state\cite{Watanuki:2018xxg}, with the results being $\Delta A_{CP} = (1.26\pm 2.4 \pm 0.67) \%$ and $\Delta_{0-} = (1.70 \pm 1.39 \pm 0.87\pm 1.15)\%$. Both results are compatible with the SM predictions and the results from \babar.

\subsection{Lepton Flavour Violation and other rare decays}

\subsubsection{Lepton Flavour Violation searches}
The lepton-flavour violating decays \decay{\Bd}{\Kstarz \mu^{\pm} e^{\mp}} and \decay{\Upsilon(3S)}{\mu^{\mp} e^{\pm} } have been searched for, setting the most stringent upper limit on the branching ratio of $1.2 - 1.8\times 10^{-7}$ at 90\% CL  for \decay{\Bd}{\Kstarz \mu^{\pm} e^{\mp}}\cite{Sandilya:2018pop}, depending on the charge combination of the leptons, and $3.6\times 10^{-7}$ at 90\% CL for \decay{\Upsilon(3S)}{\mu^{\pm} e^{\mp} }.

Limits by the \lhcb collaboration include $\BR(\decay{B_{s}^{0}}{\mu^{\pm} e^{\mp}}) < 6.3\times 10^{-9}$\cite{Aaij:2017cza},  $\BR(\decay{B^{0}}{\mu^{\pm} e^{\mp}}) < 1.3\times 10^{-9}$\cite{Aaij:2017cza} and $\BR(\decay{H}{\mu^{\pm}\tau^{\mp}}) < 26\%$\cite{Aaij:2018mea} for an SM Higgs, all at 95\% CL.

\subsubsection{Other rare decays}
Limits were set on $\BR(\decay{\Bub}{\lambda\antiproton\neu\neub}) < 3.0\times 10^{-5}$ at 90\% CL by the \babar experiment, $\BR(\decay{\Bu}{\Dsp\phi}) < 4.9\times 10^{-7}$ at 95\% CL\cite{Aaij:2017vxc}, which is a pure annihilation decay, and $\BR(\decay{\Lc}{p\mumu}) < 7.7\times 10^{-8}$ at 90\% CL\cite{Aaij:2017nsd} for a non-resonant dimuon, both by the \lhcb experiment. The last decay is dominated by decays over an $\omega$ or $\phi$ resonance. A first observation of the decay \decay{\Lc}{p\omega(\to\mumu)} was achieved, when requiring the dimuon to originate from an $\omega$ meson, yielding a branching ratio of $\BR(\decay{\Lc}{p\omega(\to\mumu})) = (9.4 \pm 3.2 \pm 1.0 \pm 2.0)\times 10^{-4}$\cite{Aaij:2017nsd}.

The decay \decay{\Bs}{\Kstarzb\mumu} is a \decay{b}{d\ell\ell} transition, strongly suppressed in the SM and therefore potentially sensitive to effects beyond the SM. An excess over background was seen for the first time by the \lhcb experiment. The integrated branching fraction was measured to be $\BR(\decay{\Bs}{\Kstarzb\mumu}) = (2.9\pm1.0\pm0.2\pm0.3)\times 10^{-8}$\cite{Aaij:2018jhg}, which corresponds to a significance of 3.4$\sigma$.

The branching fraction of the \decay{s}{d\ell\ell} transition \decay{\Sigmares^{+}}{\proton\mumu} is dominated by long-distance effects, additionally the result from the HyperCP experiment saw an excess at the lower edge of the kinematically allowed \qsq region\cite{Park:2005eka}. The branching ratio was measured to be $\BR(\decay{\Sigmares^{+}}{\proton\mumu}) = (2.2^{+1.8}_{-1.3}) \times 10^{-8}$\cite{Aaij:2017ddf} by the \lhcb experiment, which corresponds to a significance of 4.1$\sigma$. No significant structure in \qsq was observed.

%The decay \decay{\Dz}{\Km\pip \ell \ell} is a rare charm decay that was recently measured by the \lhcb collaboration in the muon channel. A first measurement of \decay{\Dz}{\Km\pip \ep \en} was performed by the \babar experiment, yielding a branching ratio of $\BR(\decay{\Dz}{\Km\pip \ep \en}) = (4.0 \pm 0.5 \pm 0.2 \pm 0.1) \times 10^{-6}$ in the region of the $\rho - \omega$ resonances. This results is compatible with the SM expectation and the \lhcb muonic result.

\section{$D$ decays}                            
\subsection{Potential for NP discovery}

Radiative charm decays provide interesting tests of New Physics, both in the up sector (directly) and as a cross check of the $B$ sector (through CKM if NP couples to weak doublets of quarks). They can be analysed in the model-independent framework of the effective Hamiltonian with three main operators contributing, $\mathcal{O}_{7,9,10}$, similarly to the $B$ physics case \cite{deBoer:2016dcg,deBoer:2017way}, and allowing for predictions for $D^0\to V\gamma$ (with $V=\rho^0, \omega,\phi, \Kstarzb$) \cite{deBoer:2017que} and $D\to \pi\ell\ell$ (branching ratio, forward-backward asymmetry, lepton-flavour universality) \cite{deBoer:2015boa,Fajfer:2015mia}. The latter are affected by very large long-distance contributions which are difficult to estimate.
They can also be used to test trendy NP models to explain $B$ anomalies, such as leptoquarks \cite{Dorsner:2016wpm,Fajfer:2015mia, Becirevic:2016oho}, with important constraints coming from $D^0\to\mu\mu$, $D_{(s)}\to \mu\nu$, $D_s\to\tau\nu$.

\subsection{Description of $D\to PP\ell\ell$}

The decay $D^0\to P_1P_2\ell^+\ell^-$ (with $P_1,P_2$ being pions and/or kaons) can provide interesting information on the $c\to u \ell\ell$ neutral current, but it requires analysing long-distance contributions~\cite{Cappiello:2012vg}. The latter can be described using a resonance saturation model, writing the decay as $D^0\to V\ell^+\ell^-$ with a subsequent decay $V\to P_1P_2$. The vector $D^0\to V\ell^+\ell^-$  can be described assuming the dominance of the lowest vector/axial resonances and factorisation on the resulting weak matrix element. Predictions can be made for various modes, separating bremsstrahlung and direct emissions, with a good overall agreement with \babar and \belle results for these modes~\cite{Aaij:2015hva,Aaij:2017iyr,Lees:2018vns}, and with other theoretical descriptions for the long-distance contributions to these modes~\cite{deBoer:2018buv}. Additional topologies involving axial (rather than vector) intermediate resonances could turn out to be important (a third of the vector contribution). The analysis could allow one to understand better the cleanliness of angular asymmetries for these modes.

\subsection{Recent experimental results}
The BESIII experiment has presented various results on rare charm meson decays. Concerning charged currents, they reported the observation of  $\BR(D^0\to a_0^- e^+\nu_e)=(1.33^{+0.33}_{-0.29}\pm 0.09)\times 10^{-4}$~\cite{Ablikim:2018ffp} and searches for $D^+\to D^0e^+\nu_e$~\cite{Ablikim:2017tdj} and $D\to\gamma e^+\nu$~\cite{Ablikim:2017twd} both constrained below $3.0\times 10^{-5}$  at 90\% CL. For neutral currents, a comprehensive list of modes $D\to he^+e^-$ and $D\to hh'e^+e^-$ (with $h$ and $h'$ light strange and/or non strange mesons) has been studied, with upper limits between $10^{-5}$ and $10^{-6}$ improving significantly compared to previous experimental bounds~\cite{Ablikim:2018gro}. Further searches are ongoing (lepton number violation through $D\to K\pi e^+e^+$, $D\to \pi^0\nu\bar\nu$).

The LHCb experiment has studied forward-backward, triple-product and \CP asymmetries in $D^0\to\pi^+\pi^-\mu^+\mu^-$ and $D^0\to K^+ K^-\mu^+\mu^-$. These modes contain both short-distance (neutral current) and long-distance (resonance) contributions. All asymmetries could be interesting probes of NP, but they show a good compatibility with zero in agreement with the Standard Model~\cite{Aaij:2018fpa}.

LHCb has also presented new results for the branching fractions of $D^+\to K^-K^+K^+$, $D^+\to \pi^-\pi^+K^+$ and 
$D_s^+\to \pi^-K^+K^+$, requiring a careful study of the background and an anlysis of the efficencies varying across the Dalitz plane. They improved significantly ratios of branching ratios between these decays, providing the world's best measurements~\cite{Aaij:2018hik}.

\section{$K$ decays}

\subsection{$\varepsilon'/\varepsilon$ in and beyond the SM}
The direct \CP violation in $K \to \pi\pi$ decays, described by the ratio $\varepsilon'/\varepsilon$, plays a very important role in the tests of the SM and more recently in the tests of its possible extensions. 
A master formula for the ratio $\varepsilon'/\varepsilon$ is obtained with a model-independent approach in the context of the $\Delta S~=~1$ effective theory with operators invariant under QCD and QED and in the context of the Standard Model Effective Field Theory (SMEFT) with the operators invariant under the full SM gauge group.
Such a formula, which allows to calculate automatically $(\varepsilon'/\varepsilon)_\text{BSM}$ once the
Wilson coefficients of all contributing operators are known at the electroweak
scale $\mu_\text{EW}$, reads as~\cite{Aebischer:2018quc}
$$ \left(\frac{\varepsilon'}{\varepsilon}\right)_\text{BSM} 
 = \sum_i P_i(\mu_\text{EW}) ~\text{Im}\left[ C_i(\mu_\text{EW})-C^\prime_i(\mu_\text{EW})\right],
$$
where
$$
 P_i(\mu_\text{EW}) = \sum_{j} \sum_{I=0,2} p_{ij}^{(I)}(\mu_\text{EW}, \mu_c)
 \,\left[\frac{\langle {\mathcal{O}}_j (\mu_c)\rangle_I}{\text{GeV}^3}\right]\,.
$$
The present master formula for $\varepsilon'/\varepsilon$ can be applied to any theory beyond the Standard Model (BSM) in which the Wilson coefficients of all contributing operators have been calculated at the electroweak scale.

\subsection{Experimental results on $K\to\pi\nu\bar{\nu}$ decays}
The $K\to\pi\nu\bar{\nu}$ decays are flavour changing neutral current processes, highly suppressed due to
quadratic GIM mechanism and to CKM suppression. The dominant contribution comes from the short-distance physics of the top quark
loop, with negligible long-distance corrections. This makes them very clean theoretically and sensitive to physics beyond the SM, probing the highest mass scales among the rare meson decays.
The NA62 experiment at CERN SPS is designed to measure the branching ratio of the $K^+\to\pi^+\nu\bar{\nu}$ decay using a novel kaon decay-in-flight technique, while the KOTO experiment at JPARC aims to study the $K_L\to\pi^0\nu\bar{\nu}$ decay.
Both experiments produced new results in 2018. The NA62
experiment observed one candidate event, which translates into an upper limit on the branching ratio~\cite{CortinaGil:2018fkc} $\mathcal{B}(K^+\to\pi^+\nu\bar{\nu}) < 14\times10^{-10}$ at  95\% CL, compatible with the Standard Model prediction. The KOTO experiment improved the
existing upper limits on the branching ratio of the neutral kaon
decay by an order of magnitude~\cite{Ahn:2018mvc}: $\mathcal{B}(K_L\to\pi^0\nu\bar{\nu}) < 3.0\times10^{-9}$ at 90\% CL.

\subsection{Lattice results on $K\to\pi\nu\bar{\nu}$ and $K\to\pi\ell^+\ell^-$ decays}
The rare $K\to\pi\nu\bar{\nu}$ and $K\to\pi\ell^+\ell^-$ decays proceed via a flavour changing neutral current and therefore may only be induced beyond tree level in the SM. This natural suppression makes these decays sensitive to the effects of potential new physics. The \CP-conserving $K\to\pi\ell^+\ell^-$ decay channels however are dominated by a single-photon exchange; this involves a sizeable long-distance hadronic contribution which represents the current major source of theoretical uncertainty. In preparation towards the computation of the long-distance contributions to these rare decay amplitudes using lattice QCD, an exploratory study using unphysical $K$ and $\pi$ masses have been performed. In particular, the $K \to \pi^+ \ell^+ \ell^-$ form factor~$V(z)$ ($z=q^2/M^2_K$) was evaluated for the first time, obtaining\cite{Christ:2016mmq} $V(z)= 1.37(36),0.68(39),0.96(64)$ for the three values of $z=-0.5594(12),-1.0530(34),-1.4653(82)$ respectively.

\subsection{Connecting $K\to\pi\nu\bar{\nu}$ and rare $B$ decays}
Lepton Flavor Universality (LFU) in the SM is ensured by the identical couplings of the electroweak gauge bosons to all three lepton flavours. This prediction has been probed at the permille level by stringent LFU tests performed in semileptonic $K$ and $\pi$ decays, in purely leptonic $\tau$ decays, and in electroweak precision observables. Recent hints of LFU violations in semileptonic $B$ decays, for both charged-current and neutral-current mediated processes, might point to BSM contributions coupled mainly to the third generation of quarks and leptons, with some small (but non-negligible) mixing with the light generations.
In order to satisfy this assumption, an Effective Field Theory (EFT) based on the $U(2)^n$ flavour symmetry is considered~\cite{Bordone:2017lsy,Bordone:2017bld}. Such a study shows that $\mathcal{O}(1)$ deviations from the SM are expected in the $K\to\pi\nu\bar{\nu}$ decays, which are the only $K$ decays involving third-generation leptons in the final state. Moreover, the correlations between $\mathcal{B}(K\to\pi\nu\bar{\nu})$ and both $\mathcal{B}(B\to\pi\nu\bar{\nu})$ and $R(D^*)$ can be exploited to distinguish between different NP scenarios.

\section{Acknowledgements}

We thank all the participants  for the quality of the presentations and the discussions during the sessions of the Working group. We also thank the organisers of the CKM18 workshop for the perfect organisation, as well as the warm and lively atmosphere of the conference.

% Using Inspire references if possible

\bibliographystyle{LHCb}
\bibliography{main}

\ifx\mcitethebibliography\mciteundefinedmacro
\PackageError{LHCb.bst}{mciteplus.sty has not been loaded}
{This bibstyle requires the use of the mciteplus package.}\fi
\providecommand{\href}[2]{#2}
\begin{mcitethebibliography}{10}
\mciteSetBstSublistMode{n}
\mciteSetBstMaxWidthForm{subitem}{\alph{mcitesubitemcount})}
\mciteSetBstSublistLabelBeginEnd{\mcitemaxwidthsubitemform\space}
{\relax}{\relax}

\bibitem{Aaij:2016flj}
LHCb, R.~Aaij {\em et~al.},
  \ifthenelse{\boolean{articletitles}}{\emph{{Measurements of the S-wave
  fraction in $B^{0}\rightarrow K^{+}\pi^{-}\mu^{+}\mu^{-}$ decays and the
  $B^{0}\rightarrow K^{\ast}(892)^{0}\mu^{+}\mu^{-}$ differential branching
  fraction}}, }{}\href{http://dx.doi.org/10.1007/JHEP11(2016)047,
  10.1007/JHEP04(2017)142}{JHEP \textbf{11} (2016) 047},
  \href{http://arxiv.org/abs/1606.04731}{{\normalfont\ttfamily
  arXiv:1606.04731}}, [Erratum: JHEP04,142(2017)]\relax
\mciteBstWouldAddEndPuncttrue
\mciteSetBstMidEndSepPunct{\mcitedefaultmidpunct}
{\mcitedefaultendpunct}{\mcitedefaultseppunct}\relax
\EndOfBibitem
\bibitem{Aaij:2014pli}
LHCb, R.~Aaij {\em et~al.},
  \ifthenelse{\boolean{articletitles}}{\emph{{Differential branching fractions
  and isospin asymmetries of $B \to K^{(*)} \mu^+ \mu^-$ decays}},
  }{}\href{http://dx.doi.org/10.1007/JHEP06(2014)133}{JHEP \textbf{06} (2014)
  133}, \href{http://arxiv.org/abs/1403.8044}{{\normalfont\ttfamily
  arXiv:1403.8044}}\relax
\mciteBstWouldAddEndPuncttrue
\mciteSetBstMidEndSepPunct{\mcitedefaultmidpunct}
{\mcitedefaultendpunct}{\mcitedefaultseppunct}\relax
\EndOfBibitem
\bibitem{Aaij:2015esa}
LHCb, R.~Aaij {\em et~al.}, \ifthenelse{\boolean{articletitles}}{\emph{{Angular
  analysis and differential branching fraction of the decay
  $B^0_s\to\phi\mu^+\mu^-$}},
  }{}\href{http://dx.doi.org/10.1007/JHEP09(2015)179}{JHEP \textbf{09} (2015)
  179}, \href{http://arxiv.org/abs/1506.08777}{{\normalfont\ttfamily
  arXiv:1506.08777}}\relax
\mciteBstWouldAddEndPuncttrue
\mciteSetBstMidEndSepPunct{\mcitedefaultmidpunct}
{\mcitedefaultendpunct}{\mcitedefaultseppunct}\relax
\EndOfBibitem
\bibitem{Aaij:2015xza}
LHCb, R.~Aaij {\em et~al.},
  \ifthenelse{\boolean{articletitles}}{\emph{{Differential branching fraction
  and angular analysis of $\Lambda^{0}_{b} \rightarrow \Lambda \mu^+\mu^-$
  decays}}, }{}\href{http://dx.doi.org/10.1007/JHEP09(2018)145,
  10.1007/JHEP06(2015)115}{JHEP \textbf{06} (2015) 115},
  \href{http://arxiv.org/abs/1503.07138}{{\normalfont\ttfamily
  arXiv:1503.07138}}, [Erratum: JHEP09,145(2018)]\relax
\mciteBstWouldAddEndPuncttrue
\mciteSetBstMidEndSepPunct{\mcitedefaultmidpunct}
{\mcitedefaultendpunct}{\mcitedefaultseppunct}\relax
\EndOfBibitem
\bibitem{Aaij:2015oid}
LHCb, R.~Aaij {\em et~al.}, \ifthenelse{\boolean{articletitles}}{\emph{{Angular
  analysis of the $B^{0} \to K^{*0} \mu^{+} \mu^{-}$ decay using 3 fb$^{-1}$ of
  integrated luminosity}},
  }{}\href{http://dx.doi.org/10.1007/JHEP02(2016)104}{JHEP \textbf{02} (2016)
  104}, \href{http://arxiv.org/abs/1512.04442}{{\normalfont\ttfamily
  arXiv:1512.04442}}\relax
\mciteBstWouldAddEndPuncttrue
\mciteSetBstMidEndSepPunct{\mcitedefaultmidpunct}
{\mcitedefaultendpunct}{\mcitedefaultseppunct}\relax
\EndOfBibitem
\bibitem{Sirunyan:2017dhj}
CMS, A.~M. Sirunyan {\em et~al.},
  \ifthenelse{\boolean{articletitles}}{\emph{{Measurement of angular parameters
  from the decay $\mathrm{B}^0 \to \mathrm{K}^{*0} \mu^+ \mu^-$ in
  proton-proton collisions at $\sqrt{s} = $ 8 TeV}},
  }{}\href{http://dx.doi.org/10.1016/j.physletb.2018.04.030}{Phys.\ Lett.\
  \textbf{B781} (2018) 517},
  \href{http://arxiv.org/abs/1710.02846}{{\normalfont\ttfamily
  arXiv:1710.02846}}\relax
\mciteBstWouldAddEndPuncttrue
\mciteSetBstMidEndSepPunct{\mcitedefaultmidpunct}
{\mcitedefaultendpunct}{\mcitedefaultseppunct}\relax
\EndOfBibitem
\bibitem{Khachatryan:2015isa}
CMS, V.~Khachatryan {\em et~al.},
  \ifthenelse{\boolean{articletitles}}{\emph{{Angular analysis of the decay
  $B^0 \to K^{*0} \mu^+ \mu^-$ from pp collisions at $\sqrt s = 8$ TeV}},
  }{}\href{http://dx.doi.org/10.1016/j.physletb.2015.12.020}{Phys.\ Lett.\
  \textbf{B753} (2016) 424},
  \href{http://arxiv.org/abs/1507.08126}{{\normalfont\ttfamily
  arXiv:1507.08126}}\relax
\mciteBstWouldAddEndPuncttrue
\mciteSetBstMidEndSepPunct{\mcitedefaultmidpunct}
{\mcitedefaultendpunct}{\mcitedefaultseppunct}\relax
\EndOfBibitem
\bibitem{Aaboud:2018krd}
ATLAS, M.~Aaboud {\em et~al.},
  \ifthenelse{\boolean{articletitles}}{\emph{{Angular analysis of $B^0_d
  \rightarrow K^{*}\mu^+\mu^-$ decays in $pp$ collisions at $\sqrt{s}= 8$ TeV
  with the ATLAS detector}},
  }{}\href{http://dx.doi.org/10.1007/JHEP10(2018)047}{JHEP \textbf{10} (2018)
  047}, \href{http://arxiv.org/abs/1805.04000}{{\normalfont\ttfamily
  arXiv:1805.04000}}\relax
\mciteBstWouldAddEndPuncttrue
\mciteSetBstMidEndSepPunct{\mcitedefaultmidpunct}
{\mcitedefaultendpunct}{\mcitedefaultseppunct}\relax
\EndOfBibitem
\bibitem{CMS:2018sus}
CMS, C.~Collaboration, \ifthenelse{\boolean{articletitles}}{\emph{{Measurement
  of angular observables in the decay B+ to K+ mu+ mu- from proton-proton
  collisions at sqrt[s]=8TeV}}, }{}\relax
\mciteBstWouldAddEndPuncttrue
\mciteSetBstMidEndSepPunct{\mcitedefaultmidpunct}
{\mcitedefaultendpunct}{\mcitedefaultseppunct}\relax
\EndOfBibitem
\bibitem{Aaij:2018gwm}
LHCb, R.~Aaij {\em et~al.}, \ifthenelse{\boolean{articletitles}}{\emph{{Angular
  moments of the decay $\Lambda_b^0 \rightarrow \Lambda \mu^{+} \mu^{-}$ at low
  hadronic recoil}}, }{}\href{http://dx.doi.org/10.1007/JHEP09(2018)146}{JHEP
  \textbf{09} (2018) 146},
  \href{http://arxiv.org/abs/1808.00264}{{\normalfont\ttfamily
  arXiv:1808.00264}}\relax
\mciteBstWouldAddEndPuncttrue
\mciteSetBstMidEndSepPunct{\mcitedefaultmidpunct}
{\mcitedefaultendpunct}{\mcitedefaultseppunct}\relax
\EndOfBibitem
\bibitem{Aaij:2016cbx}
LHCb, R.~Aaij {\em et~al.},
  \ifthenelse{\boolean{articletitles}}{\emph{{Measurement of the phase
  difference between short- and long-distance amplitudes in the $B^{+}\to
  K^{+}\mu^{+}\mu^{-}$ decay}},
  }{}\href{http://dx.doi.org/10.1140/epjc/s10052-017-4703-2}{Eur.\ Phys.\ J.\
  \textbf{C77} (2017) 161},
  \href{http://arxiv.org/abs/1612.06764}{{\normalfont\ttfamily
  arXiv:1612.06764}}\relax
\mciteBstWouldAddEndPuncttrue
\mciteSetBstMidEndSepPunct{\mcitedefaultmidpunct}
{\mcitedefaultendpunct}{\mcitedefaultseppunct}\relax
\EndOfBibitem
\bibitem{Blake:2017fyh}
T.~Blake {\em et~al.}, \ifthenelse{\boolean{articletitles}}{\emph{{An empirical
  model to determine the hadronic resonance contributions to $\overline{B}{} ^0
  \!\rightarrow \overline{K}{} ^{*0} \mu ^+ \mu ^- $ transitions}},
  }{}\href{http://dx.doi.org/10.1140/epjc/s10052-018-5937-3}{Eur.\ Phys.\ J.\
  \textbf{C78} (2018) 453},
  \href{http://arxiv.org/abs/1709.03921}{{\normalfont\ttfamily
  arXiv:1709.03921}}\relax
\mciteBstWouldAddEndPuncttrue
\mciteSetBstMidEndSepPunct{\mcitedefaultmidpunct}
{\mcitedefaultendpunct}{\mcitedefaultseppunct}\relax
\EndOfBibitem
\bibitem{Bobeth:2017vxj}
C.~Bobeth, M.~Chrzaszcz, D.~van Dyk, and J.~Virto,
  \ifthenelse{\boolean{articletitles}}{\emph{{Long-distance effects in
  $B\rightarrow K^*\ell \ell $ from analyticity}},
  }{}\href{http://dx.doi.org/10.1140/epjc/s10052-018-5918-6}{Eur.\ Phys.\ J.\
  \textbf{C78} (2018) 451},
  \href{http://arxiv.org/abs/1707.07305}{{\normalfont\ttfamily
  arXiv:1707.07305}}\relax
\mciteBstWouldAddEndPuncttrue
\mciteSetBstMidEndSepPunct{\mcitedefaultmidpunct}
{\mcitedefaultendpunct}{\mcitedefaultseppunct}\relax
\EndOfBibitem
\bibitem{Aaij:2014ora}
LHCb, R.~Aaij {\em et~al.}, \ifthenelse{\boolean{articletitles}}{\emph{{Test of
  lepton universality using $B^{+}\rightarrow K^{+}\ell^{+}\ell^{-}$ decays}},
  }{}\href{http://dx.doi.org/10.1103/PhysRevLett.113.151601}{Phys.\ Rev.\
  Lett.\  \textbf{113} (2014) 151601},
  \href{http://arxiv.org/abs/1406.6482}{{\normalfont\ttfamily
  arXiv:1406.6482}}\relax
\mciteBstWouldAddEndPuncttrue
\mciteSetBstMidEndSepPunct{\mcitedefaultmidpunct}
{\mcitedefaultendpunct}{\mcitedefaultseppunct}\relax
\EndOfBibitem
\bibitem{Aaij:2017vbb}
LHCb, R.~Aaij {\em et~al.}, \ifthenelse{\boolean{articletitles}}{\emph{{Test of
  lepton universality with $B^{0} \rightarrow K^{*0}\ell^{+}\ell^{-}$ decays}},
  }{}\href{http://dx.doi.org/10.1007/JHEP08(2017)055}{JHEP \textbf{08} (2017)
  055}, \href{http://arxiv.org/abs/1705.05802}{{\normalfont\ttfamily
  arXiv:1705.05802}}\relax
\mciteBstWouldAddEndPuncttrue
\mciteSetBstMidEndSepPunct{\mcitedefaultmidpunct}
{\mcitedefaultendpunct}{\mcitedefaultseppunct}\relax
\EndOfBibitem
\bibitem{Lees:2012tva}
BaBar, J.~P. Lees {\em et~al.},
  \ifthenelse{\boolean{articletitles}}{\emph{{Measurement of Branching
  Fractions and Rate Asymmetries in the Rare Decays $B \to K^{(*)} l^+ l^-$}},
  }{}\href{http://dx.doi.org/10.1103/PhysRevD.86.032012}{Phys.\ Rev.\
  \textbf{D86} (2012) 032012},
  \href{http://arxiv.org/abs/1204.3933}{{\normalfont\ttfamily
  arXiv:1204.3933}}\relax
\mciteBstWouldAddEndPuncttrue
\mciteSetBstMidEndSepPunct{\mcitedefaultmidpunct}
{\mcitedefaultendpunct}{\mcitedefaultseppunct}\relax
\EndOfBibitem
\bibitem{Wei:2009zv}
Belle, J.-T. Wei {\em et~al.},
  \ifthenelse{\boolean{articletitles}}{\emph{{Measurement of the Differential
  Branching Fraction and Forward-Backword Asymmetry for $B \to
  K^{(*)}\ell^+\ell^-$}},
  }{}\href{http://dx.doi.org/10.1103/PhysRevLett.103.171801}{Phys.\ Rev.\
  Lett.\  \textbf{103} (2009) 171801},
  \href{http://arxiv.org/abs/0904.0770}{{\normalfont\ttfamily
  arXiv:0904.0770}}\relax
\mciteBstWouldAddEndPuncttrue
\mciteSetBstMidEndSepPunct{\mcitedefaultmidpunct}
{\mcitedefaultendpunct}{\mcitedefaultseppunct}\relax
\EndOfBibitem
\bibitem{Aaij:2017tyk}
LHCb, R.~Aaij {\em et~al.},
  \ifthenelse{\boolean{articletitles}}{\emph{{Measurement of the ratio of
  branching fractions
  $\mathcal{B}(B_c^+\,\to\,J/\psi\tau^+\nu_\tau)$/$\mathcal{B}(B_c^+\,\to\,J/\psi\mu^+\nu_\mu)$}},
  }{}\href{http://dx.doi.org/10.1103/PhysRevLett.120.121801}{Phys.\ Rev.\
  Lett.\  \textbf{120} (2018) 121801},
  \href{http://arxiv.org/abs/1711.05623}{{\normalfont\ttfamily
  arXiv:1711.05623}}\relax
\mciteBstWouldAddEndPuncttrue
\mciteSetBstMidEndSepPunct{\mcitedefaultmidpunct}
{\mcitedefaultendpunct}{\mcitedefaultseppunct}\relax
\EndOfBibitem
\bibitem{Aaij:2015yra}
LHCb, R.~Aaij {\em et~al.},
  \ifthenelse{\boolean{articletitles}}{\emph{{Measurement of the ratio of
  branching fractions $\mathcal{B}(\bar{B}^0 \to
  D^{*+}\tau^{-}\bar{\nu}_{\tau})/\mathcal{B}(\bar{B}^0 \to
  D^{*+}\mu^{-}\bar{\nu}_{\mu})$}},
  }{}\href{http://dx.doi.org/10.1103/PhysRevLett.115.159901,
  10.1103/PhysRevLett.115.111803}{Phys.\ Rev.\ Lett.\  \textbf{115} (2015)
  111803}, \href{http://arxiv.org/abs/1506.08614}{{\normalfont\ttfamily
  arXiv:1506.08614}}, [Erratum: Phys. Rev. Lett.115,no.15,159901(2015)]\relax
\mciteBstWouldAddEndPuncttrue
\mciteSetBstMidEndSepPunct{\mcitedefaultmidpunct}
{\mcitedefaultendpunct}{\mcitedefaultseppunct}\relax
\EndOfBibitem
\bibitem{Aaij:2017uff}
LHCb, R.~Aaij {\em et~al.},
  \ifthenelse{\boolean{articletitles}}{\emph{{Measurement of the ratio of the
  $B^0 \to D^{*-} \tau^+ \nu_{\tau}$ and $B^0 \to D^{*-} \mu^+ \nu_{\mu}$
  branching fractions using three-prong $\tau$-lepton decays}},
  }{}\href{http://dx.doi.org/10.1103/PhysRevLett.120.171802}{Phys.\ Rev.\
  Lett.\  \textbf{120} (2018) 171802},
  \href{http://arxiv.org/abs/1708.08856}{{\normalfont\ttfamily
  arXiv:1708.08856}}\relax
\mciteBstWouldAddEndPuncttrue
\mciteSetBstMidEndSepPunct{\mcitedefaultmidpunct}
{\mcitedefaultendpunct}{\mcitedefaultseppunct}\relax
\EndOfBibitem
\bibitem{Lees:2012xj}
BaBar, J.~P. Lees {\em et~al.},
  \ifthenelse{\boolean{articletitles}}{\emph{{Evidence for an excess of
  $\bar{B} \to D^{(*)} \tau^-\bar{\nu}_\tau$ decays}},
  }{}\href{http://dx.doi.org/10.1103/PhysRevLett.109.101802}{Phys.\ Rev.\
  Lett.\  \textbf{109} (2012) 101802},
  \href{http://arxiv.org/abs/1205.5442}{{\normalfont\ttfamily
  arXiv:1205.5442}}\relax
\mciteBstWouldAddEndPuncttrue
\mciteSetBstMidEndSepPunct{\mcitedefaultmidpunct}
{\mcitedefaultendpunct}{\mcitedefaultseppunct}\relax
\EndOfBibitem
\bibitem{Lees:2013uzd}
BaBar, J.~P. Lees {\em et~al.},
  \ifthenelse{\boolean{articletitles}}{\emph{{Measurement of an Excess of
  $\bar{B} \to D^{(*)}\tau^- \bar{\nu}_\tau$ Decays and Implications for
  Charged Higgs Bosons}},
  }{}\href{http://dx.doi.org/10.1103/PhysRevD.88.072012}{Phys.\ Rev.\
  \textbf{D88} (2013) 072012},
  \href{http://arxiv.org/abs/1303.0571}{{\normalfont\ttfamily
  arXiv:1303.0571}}\relax
\mciteBstWouldAddEndPuncttrue
\mciteSetBstMidEndSepPunct{\mcitedefaultmidpunct}
{\mcitedefaultendpunct}{\mcitedefaultseppunct}\relax
\EndOfBibitem
\bibitem{Huschle:2015rga}
Belle, M.~Huschle {\em et~al.},
  \ifthenelse{\boolean{articletitles}}{\emph{{Measurement of the branching
  ratio of $\bar{B} \to D^{(\ast)} \tau^- \bar{\nu}_\tau$ relative to $\bar{B}
  \to D^{(\ast)} \ell^- \bar{\nu}_\ell$ decays with hadronic tagging at
  Belle}}, }{}\href{http://dx.doi.org/10.1103/PhysRevD.92.072014}{Phys.\ Rev.\
  \textbf{D92} (2015) 072014},
  \href{http://arxiv.org/abs/1507.03233}{{\normalfont\ttfamily
  arXiv:1507.03233}}\relax
\mciteBstWouldAddEndPuncttrue
\mciteSetBstMidEndSepPunct{\mcitedefaultmidpunct}
{\mcitedefaultendpunct}{\mcitedefaultseppunct}\relax
\EndOfBibitem
\bibitem{Sato:2016svk}
Belle, Y.~Sato {\em et~al.},
  \ifthenelse{\boolean{articletitles}}{\emph{{Measurement of the branching
  ratio of $\bar{B}^0 \rightarrow D^{*+} \tau^- \bar{\nu}_{\tau}$ relative to
  $\bar{B}^0 \rightarrow D^{*+} \ell^- \bar{\nu}_{\ell}$ decays with a
  semileptonic tagging method}},
  }{}\href{http://dx.doi.org/10.1103/PhysRevD.94.072007}{Phys.\ Rev.\
  \textbf{D94} (2016) 072007},
  \href{http://arxiv.org/abs/1607.07923}{{\normalfont\ttfamily
  arXiv:1607.07923}}\relax
\mciteBstWouldAddEndPuncttrue
\mciteSetBstMidEndSepPunct{\mcitedefaultmidpunct}
{\mcitedefaultendpunct}{\mcitedefaultseppunct}\relax
\EndOfBibitem
\bibitem{Hirose:2016wfn}
Belle, S.~Hirose {\em et~al.},
  \ifthenelse{\boolean{articletitles}}{\emph{{Measurement of the $\tau$ lepton
  polarization and $R(D^*)$ in the decay $\bar{B} \to D^* \tau^-
  \bar{\nu}_\tau$}},
  }{}\href{http://dx.doi.org/10.1103/PhysRevLett.118.211801}{Phys.\ Rev.\
  Lett.\  \textbf{118} (2017) 211801},
  \href{http://arxiv.org/abs/1612.00529}{{\normalfont\ttfamily
  arXiv:1612.00529}}\relax
\mciteBstWouldAddEndPuncttrue
\mciteSetBstMidEndSepPunct{\mcitedefaultmidpunct}
{\mcitedefaultendpunct}{\mcitedefaultseppunct}\relax
\EndOfBibitem
\bibitem{Hirose:2017dxl}
Belle, S.~Hirose {\em et~al.},
  \ifthenelse{\boolean{articletitles}}{\emph{{Measurement of the $\tau$ lepton
  polarization and $R(D^*)$ in the decay $\bar{B} \rightarrow D^* \tau^-
  \bar{\nu}_\tau$ with one-prong hadronic $\tau$ decays at Belle}},
  }{}\href{http://dx.doi.org/10.1103/PhysRevD.97.012004}{Phys.\ Rev.\
  \textbf{D97} (2018) 012004},
  \href{http://arxiv.org/abs/1709.00129}{{\normalfont\ttfamily
  arXiv:1709.00129}}\relax
\mciteBstWouldAddEndPuncttrue
\mciteSetBstMidEndSepPunct{\mcitedefaultmidpunct}
{\mcitedefaultendpunct}{\mcitedefaultseppunct}\relax
\EndOfBibitem
\bibitem{Aaij:2017vad}
LHCb, R.~Aaij {\em et~al.},
  \ifthenelse{\boolean{articletitles}}{\emph{{Measurement of the
  $B^0_s\to\mu^+\mu^-$ branching fraction and effective lifetime and search for
  $B^0\to\mu^+\mu^-$ decays}},
  }{}\href{http://dx.doi.org/10.1103/PhysRevLett.118.191801}{Phys.\ Rev.\
  Lett.\  \textbf{118} (2017) 191801},
  \href{http://arxiv.org/abs/1703.05747}{{\normalfont\ttfamily
  arXiv:1703.05747}}\relax
\mciteBstWouldAddEndPuncttrue
\mciteSetBstMidEndSepPunct{\mcitedefaultmidpunct}
{\mcitedefaultendpunct}{\mcitedefaultseppunct}\relax
\EndOfBibitem
\bibitem{Aaboud:2018mst}
ATLAS, M.~Aaboud {\em et~al.},
  \ifthenelse{\boolean{articletitles}}{\emph{{Study of the rare decays of
  $B^0_s$ and $B^0$ mesons into muon pairs using data collected during 2015 and
  2016 with the ATLAS detector}}, }{}Submitted to: JHEP (2018)
  \href{http://arxiv.org/abs/1812.03017}{{\normalfont\ttfamily
  arXiv:1812.03017}}\relax
\mciteBstWouldAddEndPuncttrue
\mciteSetBstMidEndSepPunct{\mcitedefaultmidpunct}
{\mcitedefaultendpunct}{\mcitedefaultseppunct}\relax
\EndOfBibitem
\bibitem{CMS:2014xfa}
CMS, LHCb, V.~Khachatryan {\em et~al.},
  \ifthenelse{\boolean{articletitles}}{\emph{{Observation of the rare
  $B^0_s\to\mu^+\mu^-$ decay from the combined analysis of CMS and LHCb data}},
  }{}\href{http://dx.doi.org/10.1038/nature14474}{Nature \textbf{522} (2015)
  68}, \href{http://arxiv.org/abs/1411.4413}{{\normalfont\ttfamily
  arXiv:1411.4413}}\relax
\mciteBstWouldAddEndPuncttrue
\mciteSetBstMidEndSepPunct{\mcitedefaultmidpunct}
{\mcitedefaultendpunct}{\mcitedefaultseppunct}\relax
\EndOfBibitem
\bibitem{deBoer:2017way}
S.~de~Boer, \ifthenelse{\boolean{articletitles}}{\emph{{Two loop virtual
  corrections to $b\rightarrow (d,s)\ell ^+\ell ^-$ and $c\rightarrow u\ell
  ^+\ell ^-$ for arbitrary momentum transfer}},
  }{}\href{http://dx.doi.org/10.1140/epjc/s10052-017-5364-x}{Eur.\ Phys.\ J.\
  \textbf{C77} (2017) 801},
  \href{http://arxiv.org/abs/1707.00988}{{\normalfont\ttfamily
  arXiv:1707.00988}}\relax
\mciteBstWouldAddEndPuncttrue
\mciteSetBstMidEndSepPunct{\mcitedefaultmidpunct}
{\mcitedefaultendpunct}{\mcitedefaultseppunct}\relax
\EndOfBibitem
\bibitem{Benzke:2017woq}
M.~Benzke, T.~Hurth, and S.~Turczyk,
  \ifthenelse{\boolean{articletitles}}{\emph{{Subleading power factorization in
  $ \bar{B}\to {X}_s{\ell}^{+}{\ell}^{-} $}},
  }{}\href{http://dx.doi.org/10.1007/JHEP10(2017)031}{JHEP \textbf{10} (2017)
  031}, \href{http://arxiv.org/abs/1705.10366}{{\normalfont\ttfamily
  arXiv:1705.10366}}\relax
\mciteBstWouldAddEndPuncttrue
\mciteSetBstMidEndSepPunct{\mcitedefaultmidpunct}
{\mcitedefaultendpunct}{\mcitedefaultseppunct}\relax
\EndOfBibitem
\bibitem{Misiak:2015xwa}
M.~Misiak {\em et~al.}, \ifthenelse{\boolean{articletitles}}{\emph{{Updated
  NNLO QCD predictions for the weak radiative B-meson decays}},
  }{}\href{http://dx.doi.org/10.1103/PhysRevLett.114.221801}{Phys.\ Rev.\
  Lett.\  \textbf{114} (2015) 221801},
  \href{http://arxiv.org/abs/1503.01789}{{\normalfont\ttfamily
  arXiv:1503.01789}}\relax
\mciteBstWouldAddEndPuncttrue
\mciteSetBstMidEndSepPunct{\mcitedefaultmidpunct}
{\mcitedefaultendpunct}{\mcitedefaultseppunct}\relax
\EndOfBibitem
\bibitem{Czakon:2015exa}
M.~Czakon {\em et~al.}, \ifthenelse{\boolean{articletitles}}{\emph{{The
  $(Q_{7}, Q_{1,2})$ contribution to $ \overline{B}\to {X}_s\gamma $ at $
  \mathcal{O}\left({\alpha}_{\mathrm{s}}^2\right) $}},
  }{}\href{http://dx.doi.org/10.1007/JHEP04(2015)168}{JHEP \textbf{04} (2015)
  168}, \href{http://arxiv.org/abs/1503.01791}{{\normalfont\ttfamily
  arXiv:1503.01791}}\relax
\mciteBstWouldAddEndPuncttrue
\mciteSetBstMidEndSepPunct{\mcitedefaultmidpunct}
{\mcitedefaultendpunct}{\mcitedefaultseppunct}\relax
\EndOfBibitem
\bibitem{Huber:2018gii}
T.~Huber, Q.~Qin, and K.~K. Vos,
  \ifthenelse{\boolean{articletitles}}{\emph{{Five-particle contributions to
  the inclusive rare $\bar B \to X_{s(d)} \, \ell^+\ell^-$ decays}},
  }{}\href{http://dx.doi.org/10.1140/epjc/s10052-018-6215-0}{Eur.\ Phys.\ J.\
  \textbf{C78} (2018) 748},
  \href{http://arxiv.org/abs/1806.11521}{{\normalfont\ttfamily
  arXiv:1806.11521}}\relax
\mciteBstWouldAddEndPuncttrue
\mciteSetBstMidEndSepPunct{\mcitedefaultmidpunct}
{\mcitedefaultendpunct}{\mcitedefaultseppunct}\relax
\EndOfBibitem
\bibitem{Altmannshofer:2017fio}
W.~Altmannshofer, C.~Niehoff, P.~Stangl, and D.~M. Straub,
  \ifthenelse{\boolean{articletitles}}{\emph{{Status of the $B\rightarrow
  K^*\mu ^+\mu ^-$ anomaly after Moriond 2017}},
  }{}\href{http://dx.doi.org/10.1140/epjc/s10052-017-4952-0}{Eur.\ Phys.\ J.\
  \textbf{C77} (2017) 377},
  \href{http://arxiv.org/abs/1703.09189}{{\normalfont\ttfamily
  arXiv:1703.09189}}\relax
\mciteBstWouldAddEndPuncttrue
\mciteSetBstMidEndSepPunct{\mcitedefaultmidpunct}
{\mcitedefaultendpunct}{\mcitedefaultseppunct}\relax
\EndOfBibitem
\bibitem{Altmannshofer:2017yso}
W.~Altmannshofer, P.~Stangl, and D.~M. Straub,
  \ifthenelse{\boolean{articletitles}}{\emph{{Interpreting Hints for Lepton
  Flavor Universality Violation}},
  }{}\href{http://dx.doi.org/10.1103/PhysRevD.96.055008}{Phys.\ Rev.\
  \textbf{D96} (2017) 055008},
  \href{http://arxiv.org/abs/1704.05435}{{\normalfont\ttfamily
  arXiv:1704.05435}}\relax
\mciteBstWouldAddEndPuncttrue
\mciteSetBstMidEndSepPunct{\mcitedefaultmidpunct}
{\mcitedefaultendpunct}{\mcitedefaultseppunct}\relax
\EndOfBibitem
\bibitem{Aaij:2014wgo}
LHCb, R.~Aaij {\em et~al.},
  \ifthenelse{\boolean{articletitles}}{\emph{{Observation of Photon
  Polarization in the b→sγ Transition}},
  }{}\href{http://dx.doi.org/10.1103/PhysRevLett.112.161801}{Phys.\ Rev.\
  Lett.\  \textbf{112} (2014) 161801},
  \href{http://arxiv.org/abs/1402.6852}{{\normalfont\ttfamily
  arXiv:1402.6852}}\relax
\mciteBstWouldAddEndPuncttrue
\mciteSetBstMidEndSepPunct{\mcitedefaultmidpunct}
{\mcitedefaultendpunct}{\mcitedefaultseppunct}\relax
\EndOfBibitem
\bibitem{HFLAV16}
Heavy Flavor Averaging Group, Y.~Amhis {\em et~al.},
  \ifthenelse{\boolean{articletitles}}{\emph{{Averages of $b$-hadron,
  $c$-hadron, and $\tau$-lepton properties as of summer 2016}},
  }{}\href{http://dx.doi.org/10.1140/epjc/s10052-017-5058-4}{Eur.\ Phys.\ J.\
  \textbf{C77} (2017) 895},
  \href{http://arxiv.org/abs/1612.07233}{{\normalfont\ttfamily
  arXiv:1612.07233}}, {updated results and plots available at
  \href{https://hflav.web.cern.ch}{{\texttt{https://hflav.web.cern.ch}}}}\relax
\mciteBstWouldAddEndPuncttrue
\mciteSetBstMidEndSepPunct{\mcitedefaultmidpunct}
{\mcitedefaultendpunct}{\mcitedefaultseppunct}\relax
\EndOfBibitem
\bibitem{Akar:2018zhv}
S.~Akar {\em et~al.}, \ifthenelse{\boolean{articletitles}}{\emph{{The
  time-dependent $CP$ asymmetry in $B^0 \to K_{\rm res} \gamma \to \pi^+ \pi^-
  K^0_{\scriptscriptstyle S} \gamma$ decays}},
  }{}\href{http://arxiv.org/abs/1802.09433}{{\normalfont\ttfamily
  arXiv:1802.09433}}\relax
\mciteBstWouldAddEndPuncttrue
\mciteSetBstMidEndSepPunct{\mcitedefaultmidpunct}
{\mcitedefaultendpunct}{\mcitedefaultseppunct}\relax
\EndOfBibitem
\bibitem{Aaij:2016ofv}
LHCb, R.~Aaij {\em et~al.}, \ifthenelse{\boolean{articletitles}}{\emph{{First
  experimental study of photon polarization in radiative $B^{0}_{s}$ decays}},
  }{}\href{http://dx.doi.org/10.1103/PhysRevLett.118.021801,
  10.1103/PhysRevLett.118.109901}{Phys.\ Rev.\ Lett.\  \textbf{118} (2017)
  021801}, \href{http://arxiv.org/abs/1609.02032}{{\normalfont\ttfamily
  arXiv:1609.02032}}, [Addendum: Phys. Rev. Lett.118,no.10,109901(2017)]\relax
\mciteBstWouldAddEndPuncttrue
\mciteSetBstMidEndSepPunct{\mcitedefaultmidpunct}
{\mcitedefaultendpunct}{\mcitedefaultseppunct}\relax
\EndOfBibitem
\bibitem{Horiguchi:2017ntw}
Belle, T.~Horiguchi {\em et~al.},
  \ifthenelse{\boolean{articletitles}}{\emph{{Evidence for Isospin Violation
  and Measurement of $CP$ Asymmetries in $B \to K^{\ast}(892) \gamma$}},
  }{}\href{http://dx.doi.org/10.1103/PhysRevLett.119.191802}{Phys.\ Rev.\
  Lett.\  \textbf{119} (2017) 191802},
  \href{http://arxiv.org/abs/1707.00394}{{\normalfont\ttfamily
  arXiv:1707.00394}}\relax
\mciteBstWouldAddEndPuncttrue
\mciteSetBstMidEndSepPunct{\mcitedefaultmidpunct}
{\mcitedefaultendpunct}{\mcitedefaultseppunct}\relax
\EndOfBibitem
\bibitem{Watanuki:2018xxg}
S.~Watanuki {\em et~al.},
  \ifthenelse{\boolean{articletitles}}{\emph{{Measurements of isospin asymmetry
  and difference of direct $CP$ asymmetries in inclusive $B \to X_s \gamma$
  decays}}, }{} in {\em {39th International Conference on High Energy Physics
  (ICHEP 2018) Seoul, Gangnam-Gu, Korea, Republic of, July 4-11, 2018}}, 2018.
\newblock \href{http://arxiv.org/abs/1807.04236}{{\normalfont\ttfamily
  arXiv:1807.04236}}\relax
\mciteBstWouldAddEndPuncttrue
\mciteSetBstMidEndSepPunct{\mcitedefaultmidpunct}
{\mcitedefaultendpunct}{\mcitedefaultseppunct}\relax
\EndOfBibitem
\bibitem{Sandilya:2018pop}
Belle, S.~Sandilya {\em et~al.},
  \ifthenelse{\boolean{articletitles}}{\emph{{Search for the
  lepton-flavor-violating decay $B^{0}\to K^{\ast 0} \mu^{\pm} e^{\mp}$}},
  }{}\href{http://dx.doi.org/10.1103/PhysRevD.98.071101}{Phys.\ Rev.\
  \textbf{D98} (2018) 071101},
  \href{http://arxiv.org/abs/1807.03267}{{\normalfont\ttfamily
  arXiv:1807.03267}}\relax
\mciteBstWouldAddEndPuncttrue
\mciteSetBstMidEndSepPunct{\mcitedefaultmidpunct}
{\mcitedefaultendpunct}{\mcitedefaultseppunct}\relax
\EndOfBibitem
\bibitem{Aaij:2017cza}
LHCb, R.~Aaij {\em et~al.}, \ifthenelse{\boolean{articletitles}}{\emph{{Search
  for the lepton-flavour violating decays B$_{(s)}^{0}
  \to e^{\pm}\mu^{\mp}$}},
  }{}\href{http://dx.doi.org/10.1007/JHEP03(2018)078}{JHEP \textbf{03} (2018)
  078}, \href{http://arxiv.org/abs/1710.04111}{{\normalfont\ttfamily
  arXiv:1710.04111}}\relax
\mciteBstWouldAddEndPuncttrue
\mciteSetBstMidEndSepPunct{\mcitedefaultmidpunct}
{\mcitedefaultendpunct}{\mcitedefaultseppunct}\relax
\EndOfBibitem
\bibitem{Aaij:2018mea}
LHCb, R.~Aaij {\em et~al.}, \ifthenelse{\boolean{articletitles}}{\emph{{Search
  for lepton-flavour-violating decays of Higgs-like bosons}},
  }{}\href{http://dx.doi.org/10.1140/epjc/s10052-018-6386-8}{Eur.\ Phys.\ J.\
  \textbf{C78} (2018) 1008},
  \href{http://arxiv.org/abs/1808.07135}{{\normalfont\ttfamily
  arXiv:1808.07135}}\relax
\mciteBstWouldAddEndPuncttrue
\mciteSetBstMidEndSepPunct{\mcitedefaultmidpunct}
{\mcitedefaultendpunct}{\mcitedefaultseppunct}\relax
\EndOfBibitem
\bibitem{Aaij:2017vxc}
LHCb, R.~Aaij {\em et~al.}, \ifthenelse{\boolean{articletitles}}{\emph{{First
  observation of $B^{+} \to D_s^{+}K^{+}K^{-}$ decays and a search for $B^{+}
  \to D_s^{+}\phi$ decays}},
  }{}\href{http://dx.doi.org/10.1007/JHEP01(2018)131}{JHEP \textbf{01} (2018)
  131}, \href{http://arxiv.org/abs/1711.05637}{{\normalfont\ttfamily
  arXiv:1711.05637}}\relax
\mciteBstWouldAddEndPuncttrue
\mciteSetBstMidEndSepPunct{\mcitedefaultmidpunct}
{\mcitedefaultendpunct}{\mcitedefaultseppunct}\relax
\EndOfBibitem
\bibitem{Aaij:2017nsd}
LHCb, R.~Aaij {\em et~al.}, \ifthenelse{\boolean{articletitles}}{\emph{{Search
  for the rare decay $\Lambda_{c}^{+} \to p\mu^+\mu^-$}},
  }{}\href{http://dx.doi.org/10.1103/PhysRevD.97.091101}{Phys.\ Rev.\
  \textbf{D97} (2018) 091101},
  \href{http://arxiv.org/abs/1712.07938}{{\normalfont\ttfamily
  arXiv:1712.07938}}\relax
\mciteBstWouldAddEndPuncttrue
\mciteSetBstMidEndSepPunct{\mcitedefaultmidpunct}
{\mcitedefaultendpunct}{\mcitedefaultseppunct}\relax
\EndOfBibitem
\bibitem{Aaij:2018jhg}
LHCb, R.~Aaij {\em et~al.},
  \ifthenelse{\boolean{articletitles}}{\emph{{Evidence for the decay $
  {B}_S^0\to {\overline{K}}^{\ast 0}{\mu}^{+}{\mu}^{-} $}},
  }{}\href{http://dx.doi.org/10.1007/JHEP07(2018)020}{JHEP \textbf{07} (2018)
  020}, \href{http://arxiv.org/abs/1804.07167}{{\normalfont\ttfamily
  arXiv:1804.07167}}\relax
\mciteBstWouldAddEndPuncttrue
\mciteSetBstMidEndSepPunct{\mcitedefaultmidpunct}
{\mcitedefaultendpunct}{\mcitedefaultseppunct}\relax
\EndOfBibitem
\bibitem{Park:2005eka}
HyperCP, H.~Park {\em et~al.},
  \ifthenelse{\boolean{articletitles}}{\emph{{Evidence for the decay Sigma+
  ---> p mu+ mu-}},
  }{}\href{http://dx.doi.org/10.1103/PhysRevLett.94.021801}{Phys.\ Rev.\ Lett.\
   \textbf{94} (2005) 021801},
  \href{http://arxiv.org/abs/hep-ex/0501014}{{\normalfont\ttfamily
  arXiv:hep-ex/0501014}}\relax
\mciteBstWouldAddEndPuncttrue
\mciteSetBstMidEndSepPunct{\mcitedefaultmidpunct}
{\mcitedefaultendpunct}{\mcitedefaultseppunct}\relax
\EndOfBibitem
\bibitem{Aaij:2017ddf}
LHCb, R.~Aaij {\em et~al.},
  \ifthenelse{\boolean{articletitles}}{\emph{{Evidence for the rare decay
  $\Sigma^+ \to p \mu^+ \mu^-$}},
  }{}\href{http://dx.doi.org/10.1103/PhysRevLett.120.221803}{Phys.\ Rev.\
  Lett.\  \textbf{120} (2018) 221803},
  \href{http://arxiv.org/abs/1712.08606}{{\normalfont\ttfamily
  arXiv:1712.08606}}\relax
\mciteBstWouldAddEndPuncttrue
\mciteSetBstMidEndSepPunct{\mcitedefaultmidpunct}
{\mcitedefaultendpunct}{\mcitedefaultseppunct}\relax
\EndOfBibitem
\bibitem{deBoer:2016dcg}
S.~de~Boer, B.~Müller, and D.~Seidel,
  \ifthenelse{\boolean{articletitles}}{\emph{{Higher-order Wilson coefficients
  for $c \to u$ transitions in the standard model}},
  }{}\href{http://dx.doi.org/10.1007/JHEP08(2016)091}{JHEP \textbf{08} (2016)
  091}, \href{http://arxiv.org/abs/1606.05521}{{\normalfont\ttfamily
  arXiv:1606.05521}}\relax
\mciteBstWouldAddEndPuncttrue
\mciteSetBstMidEndSepPunct{\mcitedefaultmidpunct}
{\mcitedefaultendpunct}{\mcitedefaultseppunct}\relax
\EndOfBibitem
\bibitem{deBoer:2017que}
S.~de~Boer and G.~Hiller, \ifthenelse{\boolean{articletitles}}{\emph{{Rare
  radiative charm decays within the standard model and beyond}},
  }{}\href{http://dx.doi.org/10.1007/JHEP08(2017)091}{JHEP \textbf{08} (2017)
  091}, \href{http://arxiv.org/abs/1701.06392}{{\normalfont\ttfamily
  arXiv:1701.06392}}\relax
\mciteBstWouldAddEndPuncttrue
\mciteSetBstMidEndSepPunct{\mcitedefaultmidpunct}
{\mcitedefaultendpunct}{\mcitedefaultseppunct}\relax
\EndOfBibitem
\bibitem{deBoer:2015boa}
S.~de~Boer and G.~Hiller, \ifthenelse{\boolean{articletitles}}{\emph{{Flavor
  and new physics opportunities with rare charm decays into leptons}},
  }{}\href{http://dx.doi.org/10.1103/PhysRevD.93.074001}{Phys.\ Rev.\
  \textbf{D93} (2016) 074001},
  \href{http://arxiv.org/abs/1510.00311}{{\normalfont\ttfamily
  arXiv:1510.00311}}\relax
\mciteBstWouldAddEndPuncttrue
\mciteSetBstMidEndSepPunct{\mcitedefaultmidpunct}
{\mcitedefaultendpunct}{\mcitedefaultseppunct}\relax
\EndOfBibitem
\bibitem{Fajfer:2015mia}
S.~Fajfer and N.~Košnik, \ifthenelse{\boolean{articletitles}}{\emph{{Prospects
  of discovering new physics in rare charm decays}},
  }{}\href{http://dx.doi.org/10.1140/epjc/s10052-015-3801-2}{Eur.\ Phys.\ J.\
  \textbf{C75} (2015) 567},
  \href{http://arxiv.org/abs/1510.00965}{{\normalfont\ttfamily
  arXiv:1510.00965}}\relax
\mciteBstWouldAddEndPuncttrue
\mciteSetBstMidEndSepPunct{\mcitedefaultmidpunct}
{\mcitedefaultendpunct}{\mcitedefaultseppunct}\relax
\EndOfBibitem
\bibitem{Dorsner:2016wpm}
I.~Doršner {\em et~al.}, \ifthenelse{\boolean{articletitles}}{\emph{{Physics
  of leptoquarks in precision experiments and at particle colliders}},
  }{}\href{http://dx.doi.org/10.1016/j.physrep.2016.06.001}{Phys.\ Rept.\
  \textbf{641} (2016) 1},
  \href{http://arxiv.org/abs/1603.04993}{{\normalfont\ttfamily
  arXiv:1603.04993}}\relax
\mciteBstWouldAddEndPuncttrue
\mciteSetBstMidEndSepPunct{\mcitedefaultmidpunct}
{\mcitedefaultendpunct}{\mcitedefaultseppunct}\relax
\EndOfBibitem
\bibitem{Becirevic:2016oho}
D.~Bečirević, N.~Košnik, O.~Sumensari, and R.~Zukanovich~Funchal,
  \ifthenelse{\boolean{articletitles}}{\emph{{Palatable Leptoquark Scenarios
  for Lepton Flavor Violation in Exclusive $b\to s\ell_1\ell_2$ modes}},
  }{}\href{http://dx.doi.org/10.1007/JHEP11(2016)035}{JHEP \textbf{11} (2016)
  035}, \href{http://arxiv.org/abs/1608.07583}{{\normalfont\ttfamily
  arXiv:1608.07583}}\relax
\mciteBstWouldAddEndPuncttrue
\mciteSetBstMidEndSepPunct{\mcitedefaultmidpunct}
{\mcitedefaultendpunct}{\mcitedefaultseppunct}\relax
\EndOfBibitem
\bibitem{Cappiello:2012vg}
L.~Cappiello, O.~Cata, and G.~D'Ambrosio,
  \ifthenelse{\boolean{articletitles}}{\emph{{Standard Model prediction and new
  physics tests for $D^0 \to h^+ h^- \ell^+ \ell^- (h=\pi,K: \ell=e,\mu)$}},
  }{}\href{http://dx.doi.org/10.1007/JHEP04(2013)135}{JHEP \textbf{04} (2013)
  135}, \href{http://arxiv.org/abs/1209.4235}{{\normalfont\ttfamily
  arXiv:1209.4235}}\relax
\mciteBstWouldAddEndPuncttrue
\mciteSetBstMidEndSepPunct{\mcitedefaultmidpunct}
{\mcitedefaultendpunct}{\mcitedefaultseppunct}\relax
\EndOfBibitem
\bibitem{Aaij:2015hva}
LHCb, R.~Aaij {\em et~al.}, \ifthenelse{\boolean{articletitles}}{\emph{{First
  observation of the decay $D^{0}\rightarrow K^{-}\pi^{+}\mu^{+}\mu^{-}$ in the
  $\rho^{0}$-$\omega$ region of the dimuon mass spectrum}},
  }{}\href{http://dx.doi.org/10.1016/j.physletb.2016.04.029}{Phys.\ Lett.\
  \textbf{B757} (2016) 558},
  \href{http://arxiv.org/abs/1510.08367}{{\normalfont\ttfamily
  arXiv:1510.08367}}\relax
\mciteBstWouldAddEndPuncttrue
\mciteSetBstMidEndSepPunct{\mcitedefaultmidpunct}
{\mcitedefaultendpunct}{\mcitedefaultseppunct}\relax
\EndOfBibitem
\bibitem{Aaij:2017iyr}
LHCb, R.~Aaij {\em et~al.},
  \ifthenelse{\boolean{articletitles}}{\emph{{Observation of $D^0$ meson decays
  to $\pi^+\pi^-\mu^+\mu^-$ and $K^+K^-\mu^+\mu^-$ final states}},
  }{}\href{http://dx.doi.org/10.1103/PhysRevLett.119.181805}{Phys.\ Rev.\
  Lett.\  \textbf{119} (2017) 181805},
  \href{http://arxiv.org/abs/1707.08377}{{\normalfont\ttfamily
  arXiv:1707.08377}}\relax
\mciteBstWouldAddEndPuncttrue
\mciteSetBstMidEndSepPunct{\mcitedefaultmidpunct}
{\mcitedefaultendpunct}{\mcitedefaultseppunct}\relax
\EndOfBibitem
\bibitem{Lees:2018vns}
BaBar, J.~P. Lees {\em et~al.},
  \ifthenelse{\boolean{articletitles}}{\emph{{Observation of the decay
  $D^0\rightarrow K^-\pi^+e^+e^-$}}, }{}Submitted to: Phys.\ Rev.\ Lett.\
  (2018) \href{http://arxiv.org/abs/1808.09680}{{\normalfont\ttfamily
  arXiv:1808.09680}}\relax
\mciteBstWouldAddEndPuncttrue
\mciteSetBstMidEndSepPunct{\mcitedefaultmidpunct}
{\mcitedefaultendpunct}{\mcitedefaultseppunct}\relax
\EndOfBibitem
\bibitem{deBoer:2018buv}
S.~De~Boer and G.~Hiller, \ifthenelse{\boolean{articletitles}}{\emph{{Null
  tests from angular distributions in $D \to P_1 P_2 l^+l^-$, $l=e,\mu$ decays
  on and off peak}},
  }{}\href{http://dx.doi.org/10.1103/PhysRevD.98.035041}{Phys.\ Rev.\
  \textbf{D98} (2018) 035041},
  \href{http://arxiv.org/abs/1805.08516}{{\normalfont\ttfamily
  arXiv:1805.08516}}\relax
\mciteBstWouldAddEndPuncttrue
\mciteSetBstMidEndSepPunct{\mcitedefaultmidpunct}
{\mcitedefaultendpunct}{\mcitedefaultseppunct}\relax
\EndOfBibitem
\bibitem{Ablikim:2018ffp}
BESIII, M.~Ablikim {\em et~al.},
  \ifthenelse{\boolean{articletitles}}{\emph{{Observation of the Semileptonic
  Decay $D^0 \to a_0(980)^- e^+ \nu_e$ and Evidence for $D^+ \to a_0(980)^0 e^+
  \nu_e$}}, }{}\href{http://dx.doi.org/10.1103/PhysRevLett.121.081802}{Phys.\
  Rev.\ Lett.\  \textbf{121} (2018) 081802},
  \href{http://arxiv.org/abs/1803.02166}{{\normalfont\ttfamily
  arXiv:1803.02166}}\relax
\mciteBstWouldAddEndPuncttrue
\mciteSetBstMidEndSepPunct{\mcitedefaultmidpunct}
{\mcitedefaultendpunct}{\mcitedefaultseppunct}\relax
\EndOfBibitem
\bibitem{Ablikim:2017tdj}
BESIII, M.~Ablikim {\em et~al.},
  \ifthenelse{\boolean{articletitles}}{\emph{{Search for the rare decay $D^+
  \to D^0 e^+\nu_e$}},
  }{}\href{http://dx.doi.org/10.1103/PhysRevD.96.092002}{Phys.\ Rev.\
  \textbf{D96} (2017) 092002},
  \href{http://arxiv.org/abs/1708.06856}{{\normalfont\ttfamily
  arXiv:1708.06856}}\relax
\mciteBstWouldAddEndPuncttrue
\mciteSetBstMidEndSepPunct{\mcitedefaultmidpunct}
{\mcitedefaultendpunct}{\mcitedefaultseppunct}\relax
\EndOfBibitem
\bibitem{Ablikim:2017twd}
BESIII, M.~Ablikim {\em et~al.},
  \ifthenelse{\boolean{articletitles}}{\emph{{Search for the radiative leptonic
  decay $D^{+}\to \gamma e^{+} {\nu}_{e}$}},
  }{}\href{http://dx.doi.org/10.1103/PhysRevD.95.071102}{Phys.\ Rev.\
  \textbf{D95} (2017) 071102},
  \href{http://arxiv.org/abs/1702.05837}{{\normalfont\ttfamily
  arXiv:1702.05837}}\relax
\mciteBstWouldAddEndPuncttrue
\mciteSetBstMidEndSepPunct{\mcitedefaultmidpunct}
{\mcitedefaultendpunct}{\mcitedefaultseppunct}\relax
\EndOfBibitem
\bibitem{Ablikim:2018gro}
BESIII, M.~Ablikim {\em et~al.},
  \ifthenelse{\boolean{articletitles}}{\emph{{Search for the rare decays $D\to
  h(h')e^+e^-$}}, }{}\href{http://dx.doi.org/10.1103/PhysRevD.97.072015}{Phys.\
  Rev.\  \textbf{D97} (2018) 072015},
  \href{http://arxiv.org/abs/1802.09752}{{\normalfont\ttfamily
  arXiv:1802.09752}}\relax
\mciteBstWouldAddEndPuncttrue
\mciteSetBstMidEndSepPunct{\mcitedefaultmidpunct}
{\mcitedefaultendpunct}{\mcitedefaultseppunct}\relax
\EndOfBibitem
\bibitem{Aaij:2018fpa}
LHCb, R.~Aaij {\em et~al.},
  \ifthenelse{\boolean{articletitles}}{\emph{{Measurement of Angular and CP
  Asymmetries in $D^0\to\pi^+\pi^-\mu^+\mu^-$ and $D^0\to K^+K^-\mu^+\mu^-$
  decays}}, }{}\href{http://dx.doi.org/10.1103/PhysRevLett.121.091801}{Phys.\
  Rev.\ Lett.\  \textbf{121} (2018) 091801},
  \href{http://arxiv.org/abs/1806.10793}{{\normalfont\ttfamily
  arXiv:1806.10793}}\relax
\mciteBstWouldAddEndPuncttrue
\mciteSetBstMidEndSepPunct{\mcitedefaultmidpunct}
{\mcitedefaultendpunct}{\mcitedefaultseppunct}\relax
\EndOfBibitem
\bibitem{Aaij:2018hik}
LHCb, R.~Aaij {\em et~al.},
  \ifthenelse{\boolean{articletitles}}{\emph{{Measurement of the branching
  fractions of the decays $D^+\rightarrow K^-K ^+K^+$, $D^+\rightarrow
  \pi^-\pi^+K^+$ and $D^+_s\rightarrow \pi^-K^+K^+$}}, }{}Submitted to: JHEP
  (2018) \href{http://arxiv.org/abs/1810.03138}{{\normalfont\ttfamily
  arXiv:1810.03138}}\relax
\mciteBstWouldAddEndPuncttrue
\mciteSetBstMidEndSepPunct{\mcitedefaultmidpunct}
{\mcitedefaultendpunct}{\mcitedefaultseppunct}\relax
\EndOfBibitem
\bibitem{Aebischer:2018quc}
J.~Aebischer {\em et~al.}, \ifthenelse{\boolean{articletitles}}{\emph{{Master
  formula for $\varepsilon'/\varepsilon$ beyond the Standard Model}},
  }{}\href{http://arxiv.org/abs/1807.02520}{{\normalfont\ttfamily
  arXiv:1807.02520}}\relax
\mciteBstWouldAddEndPuncttrue
\mciteSetBstMidEndSepPunct{\mcitedefaultmidpunct}
{\mcitedefaultendpunct}{\mcitedefaultseppunct}\relax
\EndOfBibitem
\bibitem{CortinaGil:2018fkc}
NA62, E.~Cortina~Gil {\em et~al.},
  \ifthenelse{\boolean{articletitles}}{\emph{{First search for
  $K^+\rightarrow\pi^+\nu\bar{\nu}$ using the decay-in-flight technique}},
  }{}\href{http://arxiv.org/abs/1811.08508}{{\normalfont\ttfamily
  arXiv:1811.08508}}\relax
\mciteBstWouldAddEndPuncttrue
\mciteSetBstMidEndSepPunct{\mcitedefaultmidpunct}
{\mcitedefaultendpunct}{\mcitedefaultseppunct}\relax
\EndOfBibitem
\bibitem{Ahn:2018mvc}
KOTO, J.~K. Ahn {\em et~al.},
  \ifthenelse{\boolean{articletitles}}{\emph{{Search for the $K_L \!\to\! \pi^0
  \nu \overline{\nu}$ and $K_L \!\to\! \pi^0 X^0$ decays at the J-PARC KOTO
  experiment}}, }{}\href{http://arxiv.org/abs/1810.09655}{{\normalfont\ttfamily
  arXiv:1810.09655}}\relax
\mciteBstWouldAddEndPuncttrue
\mciteSetBstMidEndSepPunct{\mcitedefaultmidpunct}
{\mcitedefaultendpunct}{\mcitedefaultseppunct}\relax
\EndOfBibitem
\bibitem{Christ:2016mmq}
N.~H. Christ {\em et~al.}, \ifthenelse{\boolean{articletitles}}{\emph{{First
  exploratory calculation of the long-distance contributions to the rare kaon
  decays $K\to\pi\ell^+\ell^-$}},
  }{}\href{http://dx.doi.org/10.1103/PhysRevD.94.114516}{Phys.\ Rev.\
  \textbf{D94} (2016) 114516},
  \href{http://arxiv.org/abs/1608.07585}{{\normalfont\ttfamily
  arXiv:1608.07585}}\relax
\mciteBstWouldAddEndPuncttrue
\mciteSetBstMidEndSepPunct{\mcitedefaultmidpunct}
{\mcitedefaultendpunct}{\mcitedefaultseppunct}\relax
\EndOfBibitem
\bibitem{Bordone:2017lsy}
M.~Bordone, D.~Buttazzo, G.~Isidori, and J.~Monnard,
  \ifthenelse{\boolean{articletitles}}{\emph{{Probing Lepton Flavour
  Universality with $K \to \pi \nu \bar\nu$ decays}},
  }{}\href{http://dx.doi.org/10.1140/epjc/s10052-017-5202-1}{Eur.\ Phys.\ J.\
  \textbf{C77} (2017) 618},
  \href{http://arxiv.org/abs/1705.10729}{{\normalfont\ttfamily
  arXiv:1705.10729}}\relax
\mciteBstWouldAddEndPuncttrue
\mciteSetBstMidEndSepPunct{\mcitedefaultmidpunct}
{\mcitedefaultendpunct}{\mcitedefaultseppunct}\relax
\EndOfBibitem
\bibitem{Bordone:2017bld}
M.~Bordone, C.~Cornella, J.~Fuentes-Martin, and G.~Isidori,
  \ifthenelse{\boolean{articletitles}}{\emph{{A three-site gauge model for
  flavor hierarchies and flavor anomalies}},
  }{}\href{http://dx.doi.org/10.1016/j.physletb.2018.02.011}{Phys.\ Lett.\
  \textbf{B779} (2018) 317},
  \href{http://arxiv.org/abs/1712.01368}{{\normalfont\ttfamily
  arXiv:1712.01368}}\relax
\mciteBstWouldAddEndPuncttrue
\mciteSetBstMidEndSepPunct{\mcitedefaultmidpunct}
{\mcitedefaultendpunct}{\mcitedefaultseppunct}\relax
\EndOfBibitem
\end{mcitethebibliography}
%\begin{thebibliography}{99}

%\end{thebibliography}

\end{document}